\documentclass[sigconf]{acmart}

\setcopyright{rightsretained}

\acmISBN{978-1-4503-9823-7}

\acmConference[ICVGIP'22]{13th Indian Conference on Computer Vision, Graphics and Image Processing}{December 2022}{Gandhinagar, India}
\acmYear{2022}
\copyrightyear{2022}

\editor{Soma Biswas}
\editor{Shanmuganathan Raman}
\editor{Amit K Roy-Chowdhury}

\acmDOI{10.1145/3571600.3571640}

\acmArticle{40}

\usepackage{svg}
\usepackage{xcolor}
\usepackage{multirow}

\usepackage{tikz}
\usetikzlibrary{positioning}
\usepackage{comment}
\usepackage{mathtools}
\usepackage{diagbox}
\usepackage{mathtools, nccmath}

\definecolor{transferColor}{HTML}{FF8000}
\definecolor{lightColor}{HTML}{00FFFF}
\definecolor{openglColor}{HTML}{5A9DE0}
\definecolor{cudaColor}{HTML}{76B900}
\definecolor{tensorColor}{HTML}{E1D5E7}
\definecolor{viewSHColor}{HTML}{FF99CC}

\copyrightyear{2022}
\acmYear{2022}
\setcopyright{acmcopyright}\acmConference[ICVGIP'22]{Proceedings of the Thirteenth Indian Conference on Computer Vision, Graphics and Image Processing}{December 8--10, 2022}{Gandhinagar, India}
\acmBooktitle{Proceedings of the Thirteenth Indian Conference on Computer Vision, Graphics and Image Processing (ICVGIP'22), December 8--10, 2022, Gandhinagar, India}
\acmPrice{15.00}
\acmDOI{10.1145/3571600.3571640}
\acmISBN{978-1-4503-9822-0/22/12}

\begin{document}

\title{Real-Time Rendering of Arbitrary Surface Geometries using Learnt Transfer}

\titlenote{\href{https://dhawal1939.github.io/projects/}{Project Page}}
\author{Sirikonda Dhawal}
\email{dhawal.sirikonda@research.iiit.ac.in}
\affiliation{%
  \institution{CVIT, KCIS, IIIT Hyderabad}
  \country{India}
  \postcode{500032}
}
\author{Aakash KT}
\email{aakash.kt@research.iiit.ac.in}
\affiliation{%
  \institution{CVIT, KCIS, IIIT Hyderabad}
  \country{India}
  \postcode{500032}
}
\author{P.J. Narayanan}
\email{pjn@iiit.ac.in}
\affiliation{%
  \institution{CVIT, KCIS, IIIT Hyderabad}
  \country{India}
  \postcode{500032}
}

\newcommand{\sloanprt}{Sloan et al.\ }
\newcommand{\mckenzie}{McKenzie et al.\ }
\newcommand{\renng}{Ng et al.\ }
\newcommand{\emphtransfer}{\emph{transfer}}
\newcommand{\emphlighting}{\emph{lighting}}

\newcommand{\aakash}[1]{\textcolor{blue}{#1}}

\newcommand{\sponza}{\textsc{Sponza} }
\newcommand{\teaset}{\textsc{Teaset} }
\newcommand{\roza}{\textsc{Roza} }
\newcommand{\dining}{\textsc{Dining-Table} }
\newcommand{\plants}{\textsc{Plants} }
\newcommand{\trimesh}{\textsc{Trimesh} }

\newcommand{\mike}{\textsc{Mike-Monster} }
\newcommand{\oldcar}{\textsc{Old-Car} }
\newcommand{\fish}{\textsc{Fish} }
\newcommand{\rabbit}{\textsc{Rabbit} }

\newcommand{\bunny}{\textsc{Stanford-Bunny} }

\newcommand{\cuda}{\textsc{CUDA} }
\newcommand{\glsl}{\textsc{GLSL} }

\newcommand{\sqboxs}{1.2ex}
\newcommand{\sqboxf}{0.6pt}
\newcommand{\sqbox}[1]{\textcolor{#1}{\rule{\sqboxs}{\sqboxs}}}
\newcommand{\sqboxEmpty}[1]{%
  \begingroup
  \setlength{\fboxrule}{\sqboxf}%
  \setlength{\fboxsep}{-\fboxrule}%
  \textcolor{#1}{\fbox{\rule{0pt}{\sqboxs}\rule{\sqboxs}{0pt}}}%
  \endgroup
}
\newcommand{\forloop}{\textsc{ForLoop} }

\newcommand{\subsubsubsection}[1]{\paragraph{#1}\mbox{}\\}
\setcounter{secnumdepth}{4}
\setcounter{tocdepth}{4}

\renewcommand{\shortauthors}{S. Dhawal, Aakash KT and P.J. Narayanan}

\begin{abstract}
Precomputed Radiance Transfer (PRT) is widely used for real-time photorealistic effects. PRT disentangles the rendering equation into {\em transfer} and {\em lighting}, enabling their precomputation. Transfer accounts for the cosine-weighted visibility of points in the scene while lighting for emitted radiance from the environment. Prior art stored precomputed transfer in a tabulated manner, either in vertex or texture space. These values are fetched with interpolation at each point for shading. Vertex space methods require densely tessellated mesh vertices for high quality images. Texture space methods require non-overlapping and area-preserving UV mapping to be available. They also require a high-resolution texture to avoid rendering artifacts. In this paper, we propose a compact {\em transfer} representation that is learnt directly on scene geometry points. Specifically, we train a small multi-layer perceptron (MLP) to predict the transfer at sampled surface points. Our approach is most beneficial where inherent mesh storage structure and natural UV mapping are not available, such as Implicit Surfaces as it learns the transfer values directly on the surface. We demonstrate real-time, photorealistic renderings of diffuse and glossy materials on SDF geometries with PRT using our approach.

\end{abstract}



\begin{teaserfigure}
  \begin{center}
     \begin{tabular}{@{} c @{\hspace{0.5mm}} c @{}}
        {
            \includegraphics[width=0.35\linewidth]{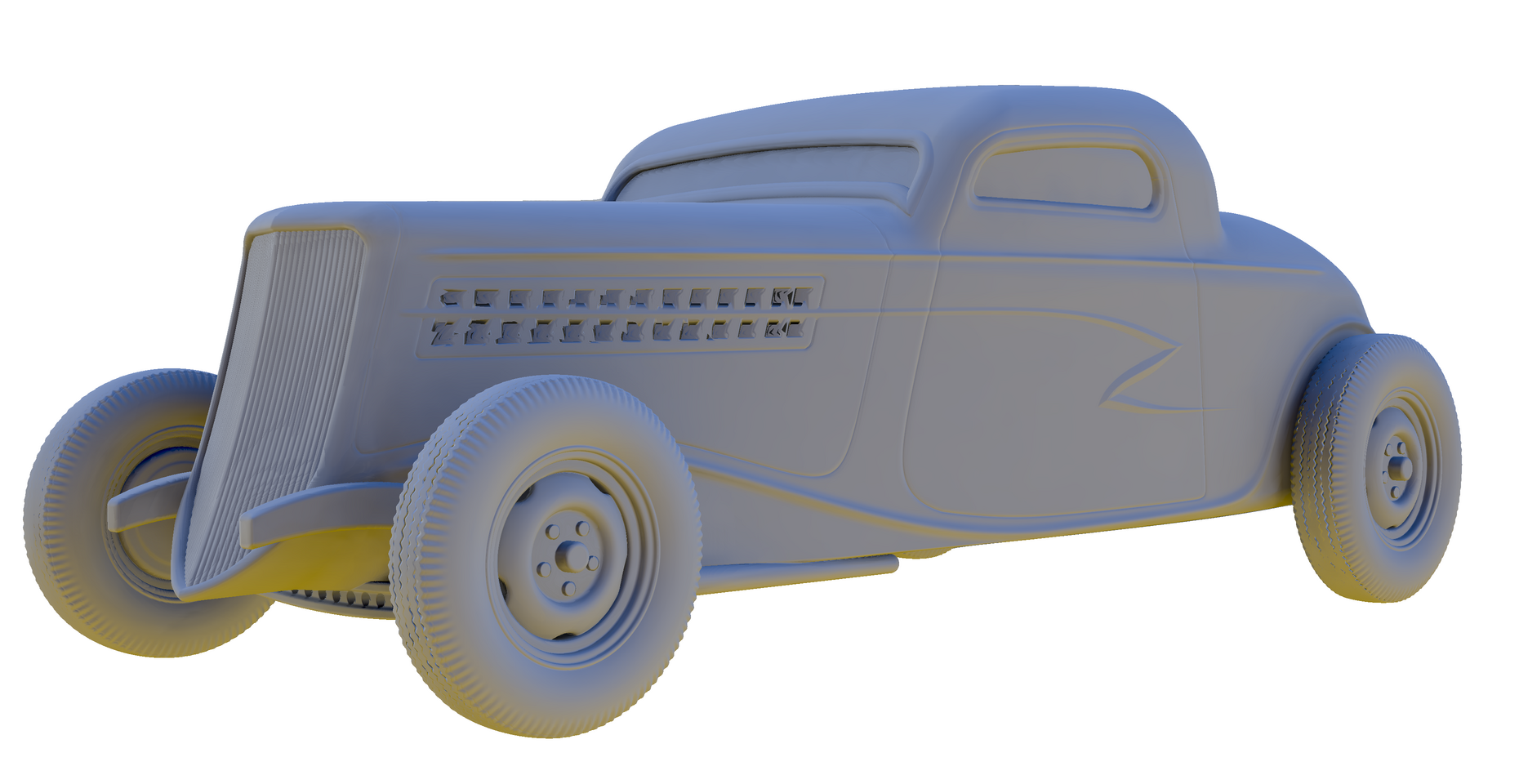}
            \label{sub_fig:diffuse_old_car}
        }
        &
        \includegraphics[width=0.35\linewidth]{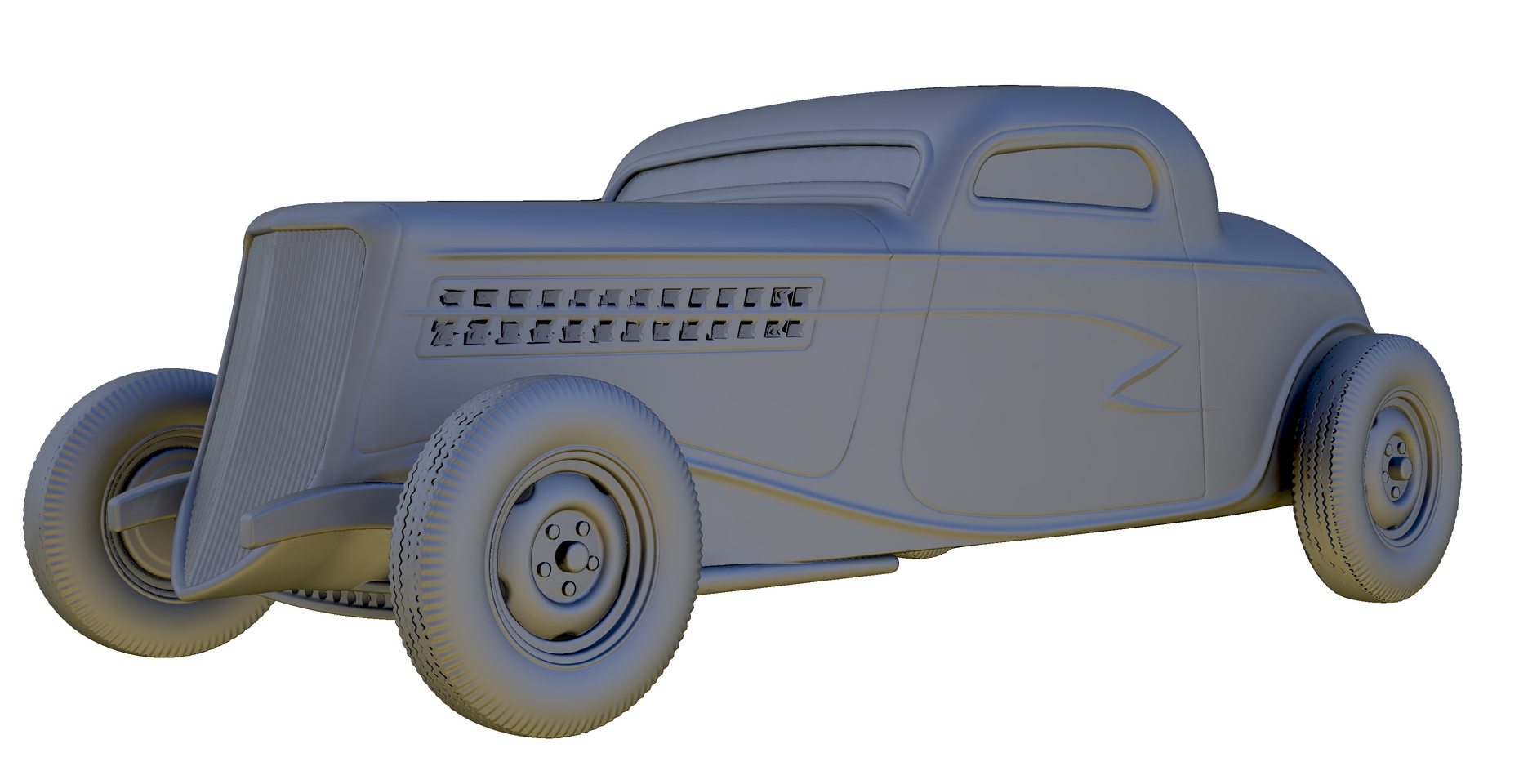} \\
        (a) & (b) \\
        \includegraphics[width=0.35\linewidth]{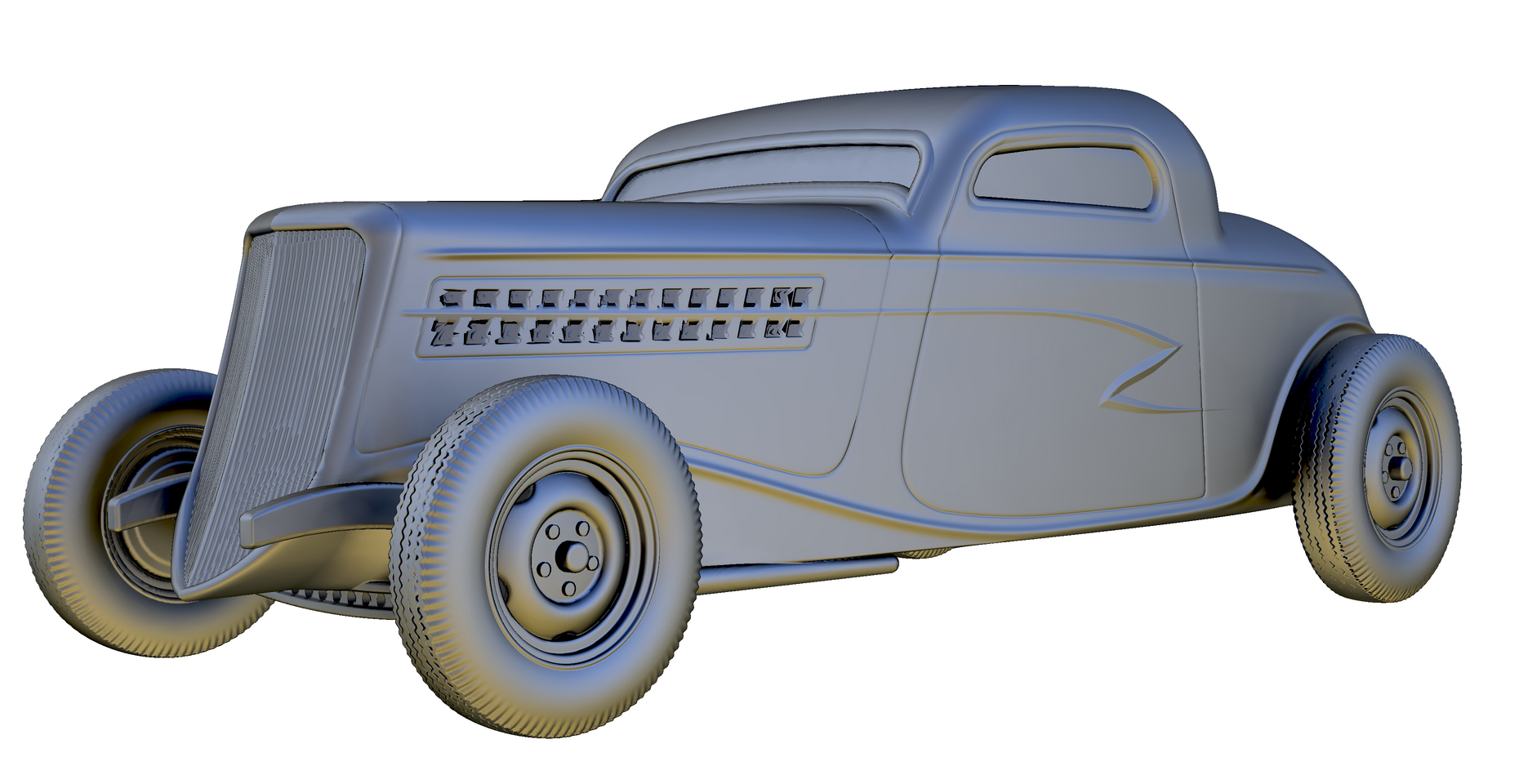} &
        \includegraphics[width=0.35\linewidth]{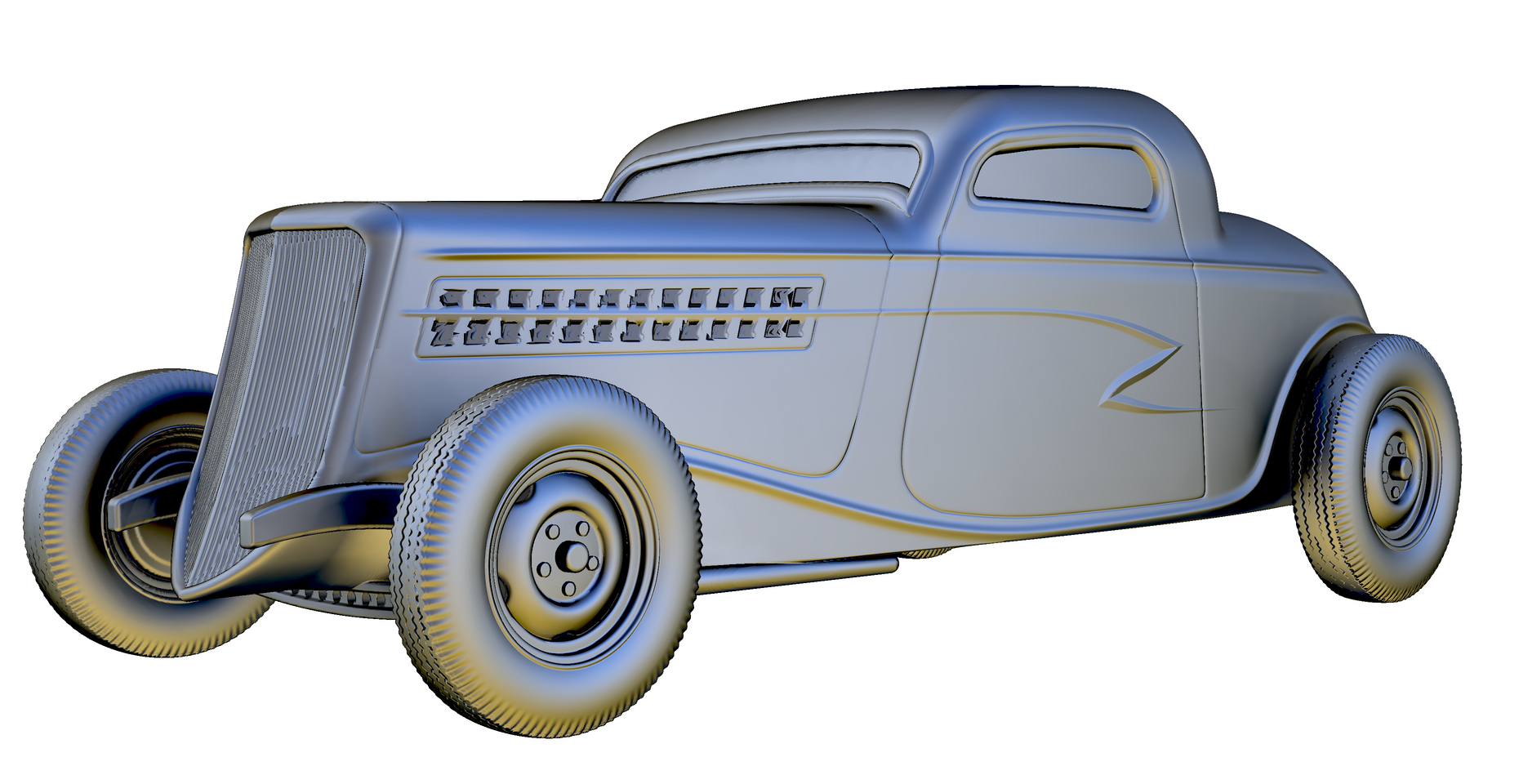} \\
        (c) & (d) \\
    \end{tabular}
  \end{center}
  \caption{We propose a learnt representation of transfer that can be used with surface representations other than meshes. In this figure, we show the \oldcar scene defined as a Signed Distance Function (SDF) rendered with our approach for two different materials types: (a) Diffuse, (b)-(d) Glossy with increasing Phong exponent. SDF rendering is not easily possible with traditional PRT methods since they rely on vertices or textures sampled with UV co-ordinates to fetch transfer, both of which are not available in SDFs. Our method plausibly renders soft shadows and glossy highlights and achieves real-time framerates.}
  \label{fig:teaser}
\end{teaserfigure}

\maketitle

\section{Introduction}
\label{sec:intro}

Photo-realistic image generation is one of the most explored problems in the field of computer graphics. Image generation with path tracing uses Monte Carlo (MC) methods to estimate the rendering equation \cite{rendering_equation}. The stochastic nature of MC introduces noise in the images which progressively reduces with additional MC samples. This makes path tracing computationally expensive for real-time applications. In Precomputed Radiance Transfer (PRT) the \textit{transfer} and \textit{lighting} of the rendering equation are calculated ahead of time and stored. The stored values are then utilized for real-time photorealistic rendering. PRT uses the Spherical Harmonic (SH) representation to efficiently store both transfer and lighting to render complex effects in video games and offline rendering for movie production \cite{prt_movie}.
    
The transfer values are spatially-varying, scene-dependent functions defined over the surface points of the scene. Conceptually, transfer values should be precomputed and stored for every surface point. At rendering time, the transfer values for each visible point should be combined with the lighting and albedo to produce the surface radiance values. Storage at all surface points is impossible in practice due to difficulty in computation, storage, and parametrization. Can we directly fit a mapping function $\sigma$ from surface parameters $(p, \hat{n})$ to transfer values $\mathcal{T}^{p}_{i}$? That is the question we address in this work. Neural approximation for complex functions is being used in many fields today. Recently, NeRF~\cite{mildenhall2020nerf} introduced radiance functions to be learned over coordinates of volumetric space. We extend this idea to fit a function $\sigma(p)$ from surface points to corresponding transfer values using a Multi-Layer Perceptron (MLP) network. The network learns a continuous representation of transfer over the surface points rather than store \textit{discrete} points. In the past, \sloanprt \cite{sloan_2002} proposed to store pre-computed transfer values at each vertex of the scene's mesh representation. The radiance computed on each vertex point was interpolated to interior points. Storing the transfer values in a texture to be interpolated during rendering  \cite{mckenzie2010textured, iwanicki, prtt} has also been tried out. Such methods need dense tessellation and/or high-resolution textures on complex scenes. These methods are limited to surface representations with good UV parametrization. Meshless hierarchical light transport \cite{lehtinen_hierarchical} removes the dependence on triangle meshes using sampled scene points and KNN search. This requires iterative deferred samplings to get good enough shading.

In this paper, we present a method that directly learns a function to generate transfer values to surface points. A small MLP holds the continuous representation of transfer values. Since the function is over surface points, our method can work on mesh models, implicit models, parametric models, etc. The MLP can directly be used for real-time rendering by encoding the network in a GLSL shader. We also present a CUDA-based system to handle larger and variable MLPs. We show high-quality rendering using PRT on surfaces represented using Signed Distance Fields (SDFs). These representations are gaining popularity \cite{single_nn_sdf, nglod} and ours is the first method of photorealistic rendering on them. We validate our renderings with traditional PRT renderings and demonstrate equivalent quality using mesh-based geometric representations Fig \ref{fig:diffuse_results}, \ref{fig:glossy_results}. Our method represents large scene by dividing them into smaller segments (Sec \ref{sub_sec:large_scenes}), each with its own MLP function.

\section{Related Work}
{
    \label{sec:related_work}
    Spherical Harmonic (SH) lighting was first proposed by \cite{irradiance_map}. This led to the development of PRT, proposed by \cite{sloan_2002, sloan2008stupid} to disentangle rendering equations into the transfer, lighting, and material while individually projecting them to the SH domain. For diffuse materials, the transfer is stored as a vector while it is stored as a matrix for glossy materials. The work by \cite{triple_ren} allows storing a vector for transfer even in-case of glossy materials by using the tripling coefficient matrix. These traditional approaches as well as the newer ones for polygonal lighting shading in PRT \cite{analytic_sh_poly_light, analytic_sh_many_poly_lights} store transfer at vertices. These approaches necessitate \textit{dense tessellation} to account for transfer changing at high frequency in the scene. Adaptive re-meshing techniques proposed by \cite{adaptivemeshsubdiv2004} store transfer at vertices while re-meshing the region where the shadows are missing. Works by \cite{iwanicki, mckenzie2010textured} store low-order SH coefficients at UV-mapped textures to account for diffuse results. The limitations of these methods are, they are constrained to mesh representation either by using a \textit{vertex storage} approach or by using \textit{a texture} storage approach. \textit{Both of these storage schemes are not present in Implicit surface representations like SDFs limiting such geometric representations to be used in PRT frameworks.}
    \par \textit{Discrete Storage of transfer:}
    {
         To remove the dependency on the mesh representation, Meshless hierarchical transport proposed by \cite{lehtinen_hierarchical} samples and stores selective points by accounting for high-fidelity changes in the transport function. But this approach requires multiple re-sampling and weighted k-nearest neighbor searches to obtain a transfer to a given query point. Another approach called Irradiance Volumes \cite{irradiance_vol} can be used to store the SH-transfer vectors in volumetric grids. Usually, this approach is prone to bleeding and leakage, causing artifacts. All of these methods rely on a discrete representation of transfer.
    }
    \par \textit{Function Representation of Transfer:}
    {
        We propose to learn a functional representation of transfer which regresses transfer given a query point in the scene geometry. While the works like \cite{rrfs} try to regress radiance, we aim to regress the transfer. In our case, the transfer is learnt on the scene geometry, which enables us to use other surface representations like implicit surface representations and parametric surface representations e.g. SDF which lack storage schemes. Recent works \cite{deepdynamic_prt} uses harmonic mapped mesh geometry to learn a transfer similar to ours to reduce the memory limitation on dynamic scene. But the work depends on the spatial information preserved between animated harmonic maps leveraging the spatial prior of convolution operations, while also still being limited to a mesh representation.  NeuralPRT by \citet{neural_prt} and Neural Radiance Transfer Fields by \citet{nrtf} take inspiration from PRT, learning latent representation for transfer with a lighting, diffuse and glossy descriptor. Furthermore, they operate in the image space and perform loss calculations in the image domain, whereas we operate in the SH space. For disentangling components from the final rendered images they employ large neural MLPs for their constituent rendering components, limiting their run times. We on the other hand stick to traditional formulations of PRT allowing our method to be easily integrated into existing frameworks like \cite{analytic_sh_many_poly_lights, analytic_sh_poly_light}
    }
}

\section{Background}
{
    \label{sec:background}
    In the Triple Product formulation \cite{triple_ren} of PRT that we use, the rendering equation for direct lighting at point $p$ is given by:
    \begin{equation}\useshortskip
        \label{eq:rendering}
        B^p(\omega_o) = \int_{\Omega} L(\omega_i) \rho^p(\omega_o, \omega_i) V^p(\omega_i) (\omega_i \odot n) d\omega_i, 
    \end{equation}
    where $\omega_o$ is the direction towards the viewer from $p$, $\omega_i$ is the incoming direction on the unit hemisphere $\Omega$ and $n$ is the surface normal at $p$. $B^p$ is the reflected radiance in direction $\omega_o$, $L$ is the incoming environment light from $\omega_i$, $V^p$ is the binary visibility function and $\rho^p$ is the Phong  Bi-directional Reflectance Distribution Function (BRDF) \cite{phong}. Eq. \ref{eq:rendering} is decomposed into the lighting $L$ and transfer $T^p(\omega_i) = V^p(\omega_i) (\omega_i \odot n)$ which are then projected to the SH basis with coefficients $\mathcal{L}_{i}$ and $\mathcal{T}^{p}_{i}$ respectively. For a glossy BRDF, the SH coefficients $\mathcal{H}^{p}_{k}$ of transferred radiance at point $p$ are given by:
    \begin{equation}\useshortskip
        \label{eq:trip_prod}
        \mathcal{H}^{p}_{k} = 
        \int_{S^2} y_k(\omega) \left ( \sum_{i=1}^{n^2} \mathcal{T}^{p}_{i} y_i(\omega) \right ) \left ( \sum_{j=1}^{n^2} \mathcal{L}_{j} y_j(\omega) \right ) d\omega = \sum_{ij} \tau_{ijk} \mathcal{T}^{p}_{i} \mathcal{L}_{j},
    \end{equation}
    where $\tau_{ijk}$ is the triple product tensor (tripling coefficient matrix) and $y$ are SH basis functions. The reflected radiance $B^p$ is calculated by convolution of $H_k^{p}$ with SH coefficients of the BRDF and evaluation at reflection direction. For a diffuse BRDF, reflected radiance $B^p$ can be directly calculated as:
    \begin{equation}\useshortskip
        \label{eq:trip_prod_1}
        B^p = \sum_i \mathcal{T}^{p}_{i} \mathcal{L}_i.
    \end{equation}
    In both cases, this formulation results in a $k$ dimensional vector for transfer at point $p$. Note that computing the reflected radiance is computationally more expensive for glossy BRDFs since it involves a matrix-vector product, followed by convolution and dot product rather than just a dot product in the case of diffuse BRDFs.
}

\begin{figure*}
        \centering
        \includegraphics[ width=0.98\linewidth]{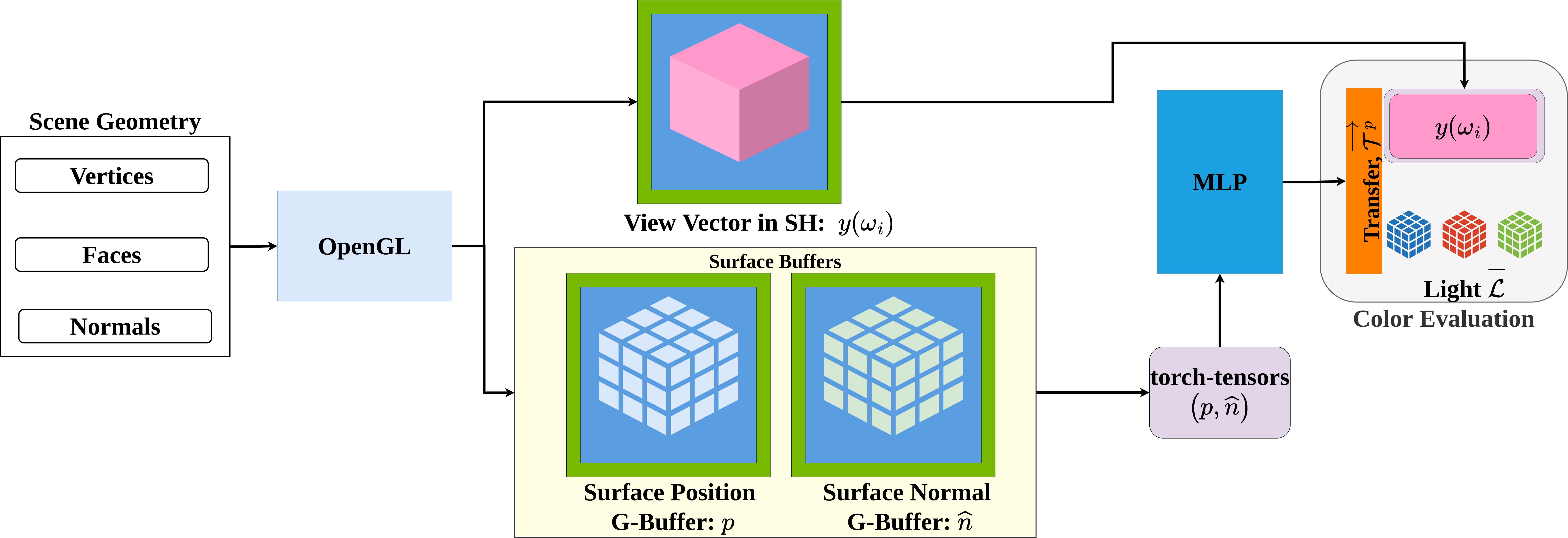}
        \caption{\cuda Implementation: We extract G-buffers that store the position $p$ and normal $\hat{n}$ in texture space. Along with G-buffers we also project per fragment view directions to SH and store them in textures. All OpenGL Textures (\sqbox{openglColor}), are shared with \cuda buffers(\sqbox{cudaColor}). The \cuda buffers are mapped to the same memory location as OpenGL texture to avoid expensive GPU - Host transfers. The \cuda buffers are copied to preloaded torch-sensors(\sqbox{tensorColor}) residing on GPU. The tensors are then fed to the network residing on GPU and the output is obtained. Since the operations are only between GPU-GPU without involving the host we avoid host latency. The network outputs a transfer vector which is clubbed with lighting and SH representation of view direction extracted as torch tensor to obtain color at each pixel using the Triple Product Formulation \cite{triple_ren}. Note: Before passing the G-buffers to MLP, fragments that intersect the geometry are separated from the ones that do not and packed to avoid unnecessary network computations.}
        \label{fig:cuda_implementation}\vspace*{-3mm}
\end{figure*}
    
\section{Method}
{
    \label{sec:method}
    In this section, we first motivate and describe our learnt representation of transfer. Next, we discuss training details along with training data generation which ensures that our learnt representation is continuous. Finally, we describe the GLSL and CUDA implementations of our learnt representation. We evaluate and show the results of both these implementations in Sect. \ref{sec:validation_results}.
    \begin{figure}[H]
        \centering
        \includegraphics[width=0.5\linewidth]{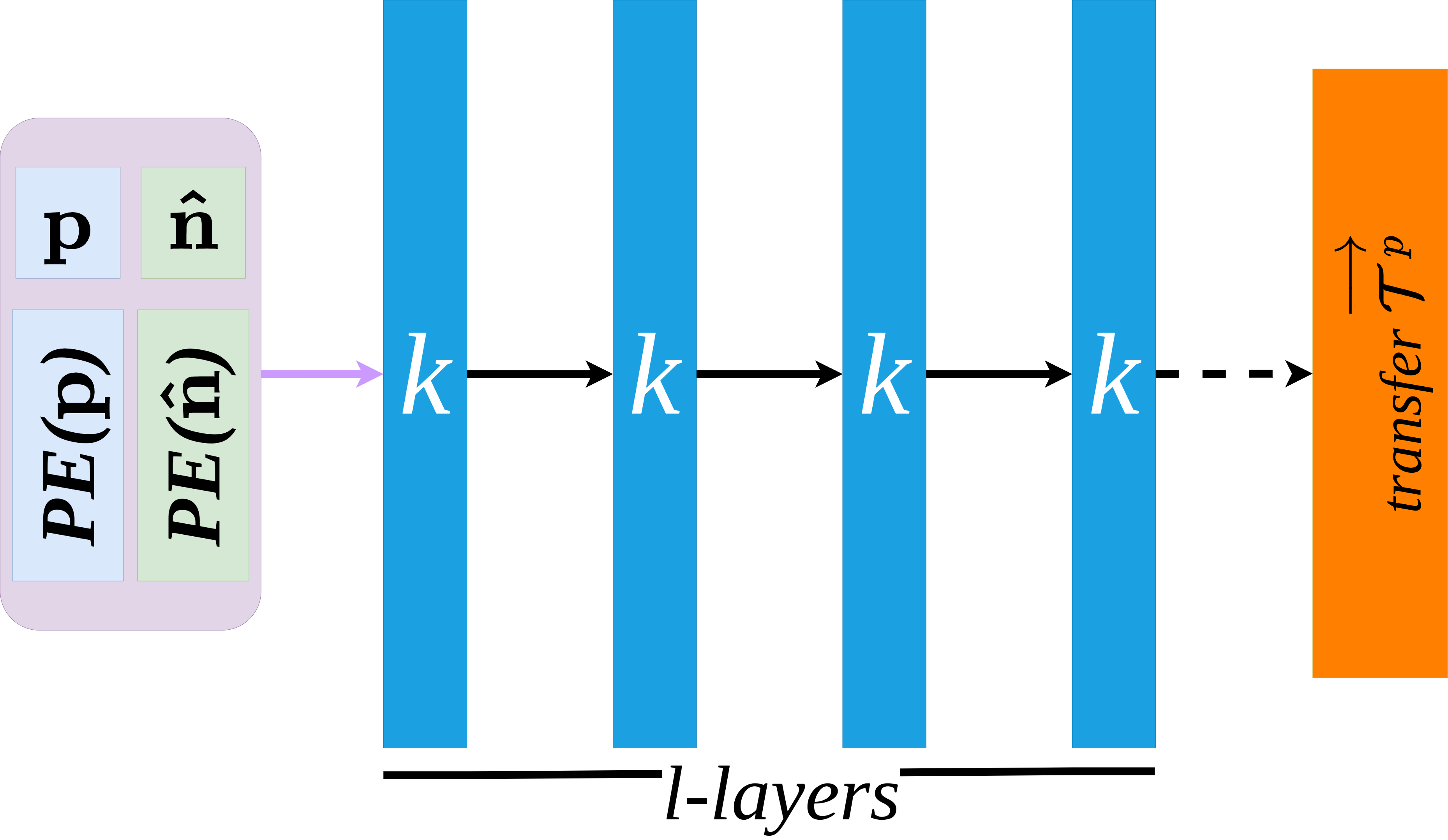}
        \caption{Our network is a Multi-Layer Perceptron (MLP) with $k$ neurons per layer and $l$ layers. A scene point $p$ and its corresponding normal ($\hat{n}$) are fed to the network as input with positional encoding. The network outputs a vector, which are SH coefficients of {\em transfer}. The black-arrow lines represent a \textit{leaky-relu} activation while the dotted arrow represents a \textit{tanh} activation. Scenes rendered in this paper use $k=64, l=4$ or $k=128, l=4$ unless otherwise specified.}
        \label{fig:mlp_figure}\vspace*{-3mm}
    \end{figure}
    \subsection{A learnt representation of transfer}
     {
        \label{sub_sec:learnt_transfer}
    
        The PRT formulation in Equation \ref{eq:trip_prod_1} describes the evaluation of direct lighting at every point $p$. Ideally, it needs to be pre-computed at every visible point in the rendered view requiring {transfer-vectors} ($\overrightarrow{\mathcal{T}^p}$) at all surface points of geometry. This is practically impossible. Instead, we learn a continuous function on the surface of the scene geometry as a function $\sigma$
            \begin{equation}
                \label{eq:surface_mapping}
                \sigma: (p, \hat{n}) \to \overrightarrow{\mathcal{T}^p}
            \end{equation}
        that maps surface position($p$) and normal($\hat{n}$) to a transfer value. We use a Multi-Layered Perceptron (MLP) network as $\sigma$. Neural networks are known to be universal function approximators \cite{universal_funct_approx}. Recently, neural representations using coordinate-base MLPs have found a great utility to learn the distribution of radiance in a scene \cite{mildenhall2020nerf, gafani2021, Reiser2021ICCV, nerfactor2021}. We adapt this idea and learn $\sigma$ as an MLP to represent a continuous function over the given scene surface. We train the MLP by precomputing the transfer values at densely sampled surface points. This method works on all types of surface representations: meshes, implicit surfaces, SDFs, parametric surfaces, etc. 
        
        Formally, given a scene point $p$ and its normal $\hat{n}$, we train a MLP to learn a mapping function $\sigma$ presented in Eq \ref{eq:surface_mapping} to learn the SH-vector of transfer $\overrightarrow{\mathcal{T}^p}$ with components $\mathcal{T}^p_i$ (Eq. \ref{eq:trip_prod}, \ref{eq:trip_prod_1}). In our experiments, we use $16$ SH coefficients similar to \citet{analytic_sh_many_poly_lights}, making $\overrightarrow{\mathcal{T}^p}$ $16$ dimensional. The network architecture used is shown in Fig \ref{fig:mlp_figure}. It contains $l$ layers with $k$ neurons per layer. Each layer is followed by the {leaky-relu} \cite{Maas2013RectifierNI} activation function, except the last which is followed by \textit{tanh} activation. In practice, the network inputs are projected to a high dimensional representation using positional encoding \cite{pmlr-v97-rahaman19a, mildenhall2020nerf, tancik2020fourfeat}. We set $k=64$ and $l=4$ or $k=128$ and $l=4$ for scenes in this paper unless otherwise specified.
        Previously, \citet{sloan_2002} and \citet{adaptivemeshsubdiv2004} stored transfer values on mesh vertices. Textures in a \textit{UV-space} were also used to store transfer values \cite{mckenzie2010textured,prtt}. These strategies strongly rely on dictionary storage structures inherently present in mesh-based geometric representations and do not extend to implicitly represented scene objects. They also incur high memory costs to store transfers for complex scenes.
    }
    \subsection{Training details}
    {
        \label{sub_sec:training_details}
        
        Our training dataset consists of transfer values for densely sampled points in the scene. We ensure an equal distribution of these points by an area-based sampling of the mesh surface \cite{turk1990}. The ground truth transfer is calculated using ray-tracing for each sampled point, as is done traditionally \cite{prt_ravir}. Note that training data can be generated for any other geometric representation by first converting it to a mesh and then applying the above routine. For example, in the SDF case, we use marching cubes to first extract a mesh, densely sample the mesh using the above method and project these points to the SDF surface. The Transfer is then calculated for these points by sphere tracing the SDF. We project the resulting transfer to SH basis with $16$ coefficients which are used as ground truth. By training our network with dense and equally distributed samples, we ensure that it learns a continuous representation of transfer.
        For scenes used in this paper, we generate $800k - 1M$ points on the geometric surface and calculate the associated transfer vectors to train our network. The data generation takes around 2 hours per scene. We use a batch size of 8192 and train our network with this data for around 200 epochs with the $\ell_1$ loss. The time taken for training is about 10 minutes on NVIDIA RTX 3090 GPU. We use sampled points as a training set and try to regress on the vertex locations which are not included in the training sample to ensure the fitted function is consistent with the surface geometry. Our sampling scheme \cite{turk1990} ensures that the sampled points do not co-inside with the vertex locations whereby keeping the test set different from the train set ensuring uniform fitting rather than an overfit of the region. Refer to Suppl. for more details about training and sampling.
    }    
    \subsection{Real-time rendering with GLSL \& CUDA}
    {
        \label{sub_sec:rendering_implementation}
        
        Once trained on a given scene, our network can output transfer vectors for \textit{any} point in that scene.  Furthermore, since our network is small, it allows for efficient per-pixel evaluation on the GPU. Below we discuss two different implementations, one with GLSL and the other with CUDA.
        
        \subsubsection{GLSL}
        {
            \label{sub_sec:glsl_implementation}
            We implement the GLSL version within the ModernGL framework \cite{moderngl}. The network weights and biases are hardcoded in $4\times4$ matrices (\textbf{mat4} type) in the fragment shader. Specifically, the weights of each layer are divided to fit into multiple \textbf{mat4}'s. This allows for an efficient forward pass using matrix-vector products on the GPU. We have automated this using a script-based extraction of weights into the shader. With the network weights available as matrices within the shader, the rendering proceeds as follows. We first obtain the surface position $p$, its normal $\hat{n}$, and the view vector $\omega_o$ in the fragment shader using the standard OpenGL pipeline. An early depth pass is used to avoid the processing of unnecessary fragments. Next, we apply positional encoding \cite{pmlr-v97-rahaman19a} to $p$ and $\hat{n}$ and evaluate the forward pass of our network with the hardcoded weights. This results in the SH vector of transfer $\overrightarrow{\mathcal{T}^p}$ at $p$. The transfer is used with the triple product formulation \cite{triple_ren} with a global light matrix to obtain the shading at $p$ according to Eq. \ref{eq:trip_prod}, \ref{eq:trip_prod_1}. Our GLSL implementation works for both AMD and NVIDIA GPUs since OpenGL itself is supported on both hardware architectures. A downside of this implementation is that the network size is constrained by the amount of local memory available for each shader.
        }
        
        \subsubsection{CUDA} 
        {
            \label{subsub_sec:cuda_impl}
            To alleviate the dependence of network size on local shader memory, we implement the second version in CUDA. Before rendering starts, we pre-load the trained weights of our network on the GPU. Rendering is then performed in two separate passes: (1) An OpenGL pass to extract G-buffers, (2) A dedicated CUDA pass for network forward evaluation using the pre-loaded weights. The OpenGL pass works similarly to the GLSL implementation, except it only extracts G-buffers and does not perform shading. We create CUDA buffers that point to the resulting G-buffers from the previous step, which implies that these share the same GPU memory. These CUDA buffers are then converted to PyTorch \cite{pytorch} tensors using the PyCuda \cite{kloeckner_pycuda_2012} interface. This approach avoids expensive transfers between GPU and the host and is crucial to achieving real-time framerates. The PyTorch tensors are then used for the network forward pass with positional encoding. The network evaluation only processes parts of the G-buffers that intersect geometry by packing them before the network forward evaluation. The transfer vectors obtained from the network are then unpacked to their original locations in the G-buffer. This avoids unnecessary evaluations of the network further improving the FPS. The final image is obtained by combining this transfer with global light matrices with the triple product formulation \cite{triple_ren}, similar to the GLSL pass.  This implementation allows the network to be as large as the entire GPU memory since it has a dedicated forward pass for it. Refer to Fig.\ref{fig:cuda_implementation} for a visual explanation of this implementation and a caption for details.
        }
        {\color{black}
        \subsection{Handling Large Scene without loss of performance}
        {
            \label{sub_sec:large_scenes}
            In certain scenarios, while handling large scenes, a small neural network might not be sufficient to regress the \textit{transfer}-accounting for the cosine weighted visibility of the rendering equation. In such a scenario, we can either engage a large neural network or subdivide the scene into smaller segments and assign a small network to regress the respective transfer values. The latter is a \textit{more performance-friendly} approach as the number of operations required to calculate transfer will be substantially smaller than the former. This is due to the fact that per fragment only a small MLP needs to be evaluated as opposed to a large network, resulting in real-time performance obtaining render-framerates of over ~200FPS.
            To achieve this we subdivided the scene into sub-scenes and sample points in each sub-scene using the strategy explained in Sec.\ref{sub_sec:training_details}. To maintain continuous seamless regression of transfer at boundaries we sample points for extra $\delta$ area over the adjacent sub-scene refer to Fig.\ref{fig:overlap} for details. For each sample point, we trace rays into the un-subdivided(whole scene) scene. This ensures the visibility accounting for the full scene rather than the sub-scene.  Once transfer vectors are calculated, to ensure a minimal number of MLPs being used to regress the transfer of the whole scene we cluster neighboring sub-scenes using Variances analysis of the transfer vectors similar to the techniques of CPCA\cite{pca_prt_sloan}. Contrary to their method, we do not use Principle components rather just rely on the small MLPs to regress transfer. At run-time, every sub-scene is assigned respective MLP either in \glsl or \cuda based implementation. Thus, enabling parallel execution of the fragments achieving a real-time framerates of ~200FPS. A visual representation of the above strategy is given in Fig. \ref{fig:large_scenes}.
            \begin{figure}
            \centering
            \includegraphics[width=\linewidth]{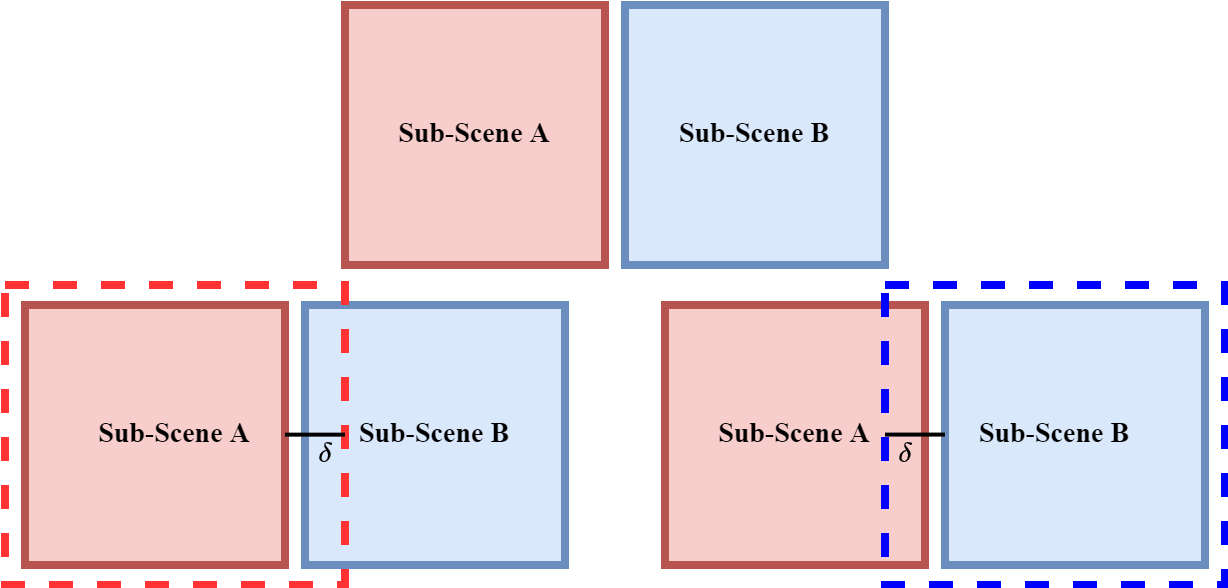}
            \caption{Sampling at boundaries: While handling the boundary regions of neighboring sub-scenes which are caused due to the sub-division. We sample $\delta$ area more into the adjacent sub-scene and include the sample points falling into that $\delta$ region into the training set of the sub-scenes MLP. This ensures smooth learning of transfer vectors and avoids seam artifacts.}
            \label{fig:overlap}\vspace*{-3mm}
            \end{figure}
            
            \begin{figure}
            \centering
            \includegraphics[width=\linewidth]{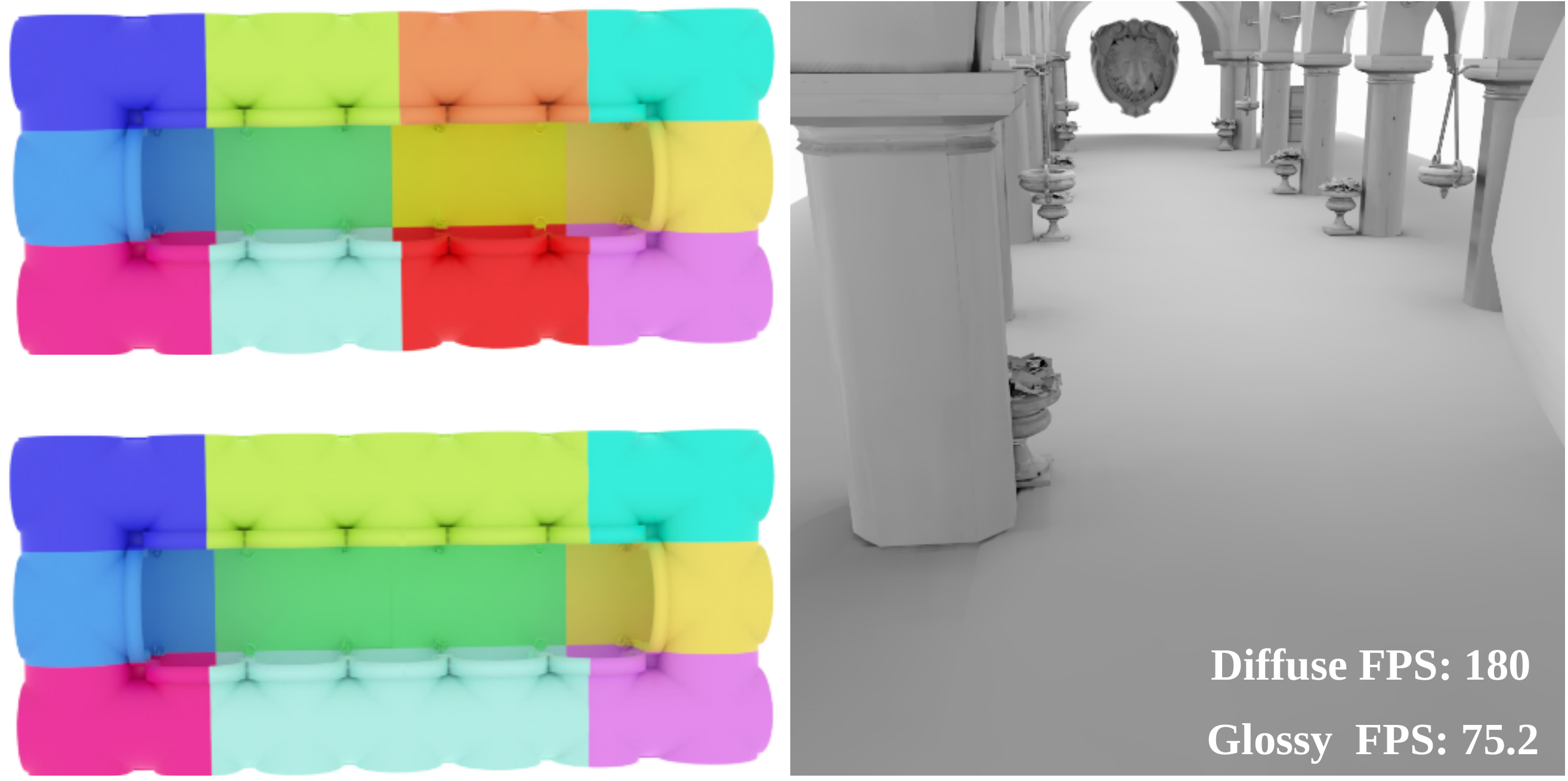}
            \caption{Large Scene: The Figure on the \textit{top-left} shows the top-view of \sponza Cathedral divided into 12 even parts. We use the heuristic presented in Sec.\ref{sub_sec:large_scenes} to join the few sub-scene to reduce the number of networks required to regress the Transfer. The clubbed geometries are visualized in the \textit{bottom-left} image where the components of the middle section are joined with their respective neighbors. Please observe that both \textit{top-left} and \textit{bottom-left} images are color-coded, each unique colors represent a sub-scene. We utilize the \textit{bottom-left} configuration of the sub-scene which only requires 9 small MLPs rather than 12 as in the case of \textit{top-left}. We visualized the resulting render in the \textit{right}. It is to be noted that we have used White light to show the transfer regressed from the network is independent of the Lighting and material of the object and can be swapped with ease as depicted in Fig.(\ref{fig:analytical_sdf_glossy_results},\ref{fig:teaser}). FPS of respective materials are mentioned in Figure.}
            \label{fig:large_scenes}\vspace*{-3mm}
            \end{figure}
        }
        }
}

\section{Validation \& Results}\label{sec:validation_results}
{
    In this section, we validate and show the results of our learnt representation (Sect. \ref{sub_sec:learnt_transfer}) implemented in GLSL and CUDA (Sect. \ref{sub_sec:rendering_implementation}). Our network is trained for each scene and we generate training data as outlined in Sect. \ref{sub_sec:training_details}. We compare our renderings with baseline PRT, which uses texture-based storage of transfer \cite{prtt, mckenzie2010textured} with the triple product formulation \cite{triple_ren}. We use the texture version as the baseline since it produces artifact-free renderings in most cases and avoids the memory overhead of vertex-based approaches for dense meshes. The baseline implementation stores transfer in a $1024\times1024$ texture with optional texture-sets for large meshes. In Sect. \ref{sub_sec:validation_triangle_mesh}, we validate our renderings of triangle meshes by qualitative and quantitative comparison with baseline PRT. In Sect. \ref{sub_sec:analytical_sdf_results}, we show results on scenes with SDF as the geometric representation. This is possible since our network learns a continuous representation of transfer and can predict it for any point in the scene. We show results on both analytical and neural SDFs. Lastly in Sect. \ref{sub_sec:cuda_vs_glsl}, we compare and contrast our GLSL and CUDA implementations and discuss their framerates for different network sizes. All scenes are rendered using our GLSL implementation (unless otherwise specified) on an NVIDIA RTX 3090 GPU and at a resolution of $1024\times1024$.
    \begin{figure}
        \centering
        \includegraphics[width=\linewidth]{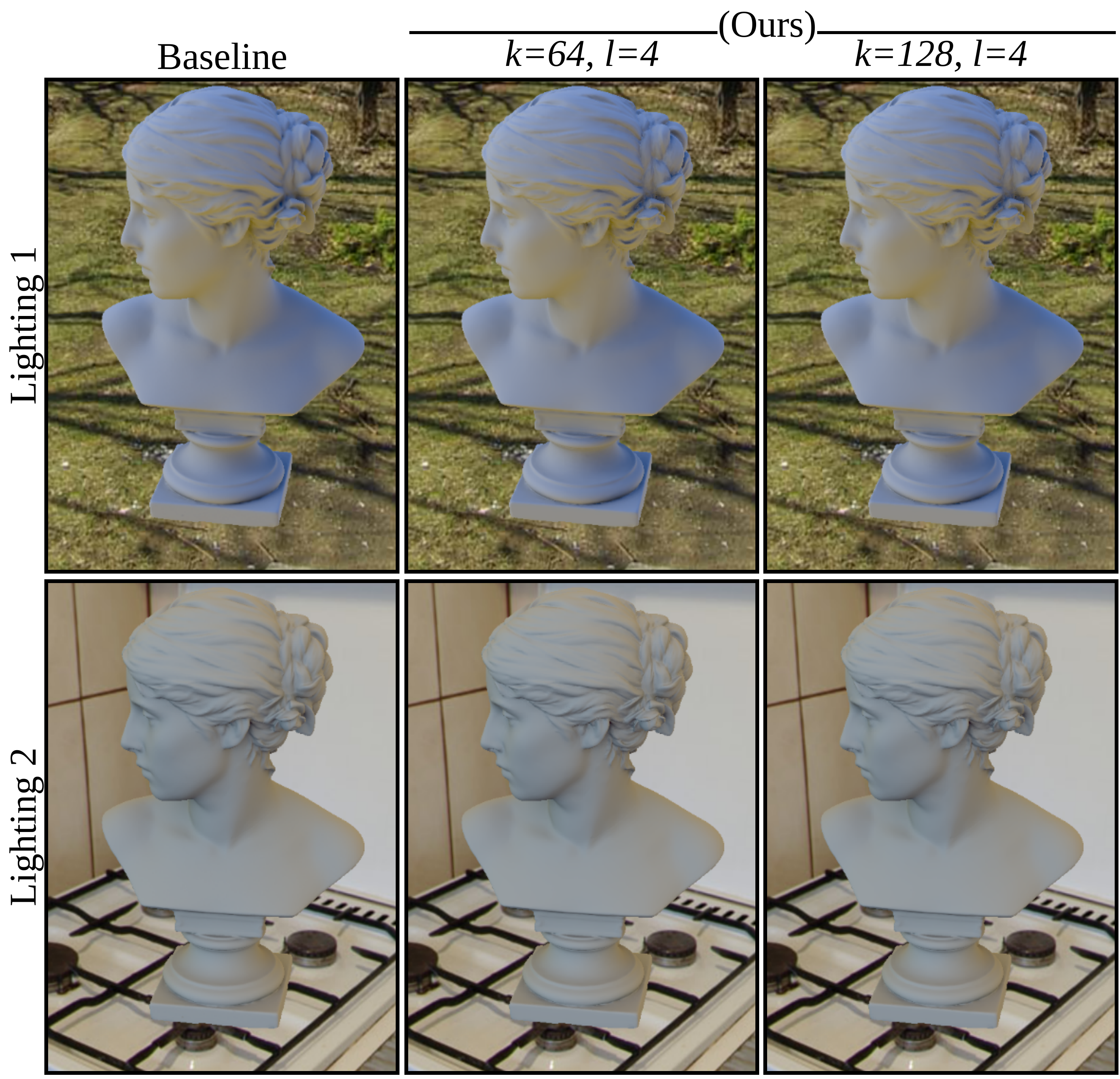}
        \caption{Diffuse Mesh Results: Since our learned function is only accounting for transfer we can dynamically change the light on the fly without retraining the network. We show results of \roza with two different lightings while using the same set of learned weights. Our approach plausibly renders soft shadows, especially in high-frequency visibility changes in the hair.}
        \label{fig:diffuse_results}\vspace*{-3mm}
    \end{figure}
    \begin{figure*}
        \centering
        \includegraphics[width=\linewidth]{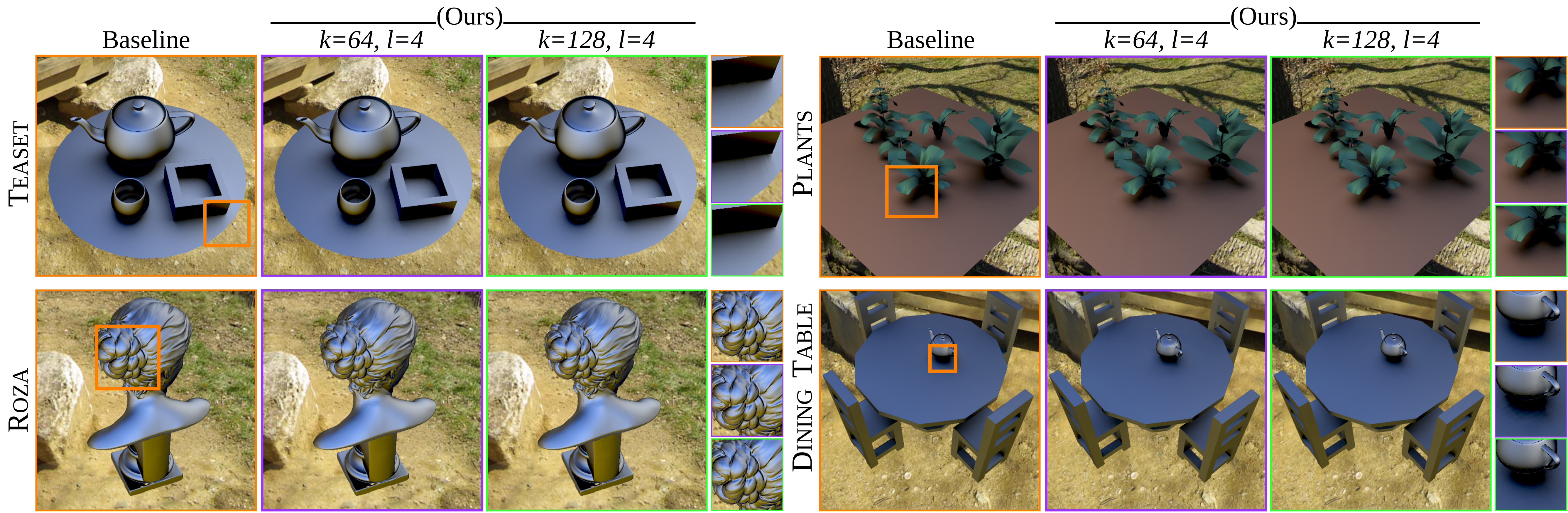}
        \caption{We show the results of our approach and compare them with the baseline (left), which is a texture-based transfer storage implementation of PRT. Our results are rendered with $k=64, l=4$ network (middle) and $k=128, l=4$ network (right). Insets are provided at the right-extreme, where we can observe that $k=64, l=4$ network produces plausible results albeit with a few jaded lines in the case of \plants and \dining, while $k=128, l=4$ network gets rid of those artifacts and matches the baseline. We note that $k=64, l=4$ network comfortably achieves real-time FPS while $k=128, l=4$ has a lower FPS (Tbl. \ref{tab:mesh_network_comparisions}). \textit{Please inspect insets closely}} 
        \label{fig:glossy_results}\vspace*{-3mm}
    \end{figure*}
    \subsection{Validation on triangle meshes}
        \label{sub_sec:validation_triangle_mesh}
        
        We validate our results on four scenes with triangle meshes: \teaset, \plants, \roza, \& \dining. {\color{black}Additionally, we also show results on a large scene leveraging the strategy presented in Sec.~\ref{sub_sec:large_scenes} and Fig.~\ref{fig:large_scenes} shows rendered results.} These scenes contain high-frequency changes in visibility, for example in the hair of \roza and shadows due to leaves in \plants. And our network is able to regress the SH vectors of transfer with ease. We render our results with diffuse and glossy materials using two different sizes of the network: $k=64, l=4$ and $k=128, l=4$ and compare them with baseline PRT. Rendering results are shown in Fig. \ref{fig:glossy_results} and quantitative metrics (MAE, PSNR, SSIM) along with frame times (FPS) are shown in Table \ref{tab:mesh_network_comparisions}. The quantitative metrics are calculated by comparing our rendering with baseline PRT as a reference. Our rendering results closely match the baseline and achieve high PSNR and SSIM values, which validates our method. We further compare our rendering of the diffuse \roza scene with baseline PRT in Fig. \ref{fig:diffuse_results}. Our method is able to reproduce fine soft shadows which further strengthens our validation. As shown in Table \ref{tab:mesh_network_comparisions}, the $k=64, l=4$ network comfortably achieves real-time FPS for both diffuse and glossy materials. A larger network ($k=128, l=4$) achieves better rendering quality and metrics albeit at a drop in FPS. 
        
        \begin{table*}
            \centering
            \begin{tabular}{|c | c | c | c | c | c | c |c |c|}
                \hline
                \multicolumn{9}{|c|}{Diffuse}\\
                \hline
                \multirow{2}{*}{Scene} & \multicolumn{4}{c|}{\textit{k=64, l=4}} & \multicolumn{4}{c|}{\textit{k=128, l=4}}\\
                \cline{2-9}
                        & MAE     & PSNR   & SSIM    & FPS   & MAE     & PSNR   & SSIM    & FPS  \\
                \hline
                \teaset & 0.00074 & 49.227 & 0.99708 & 350.5 & 0.00055 & 51.050 & 0.99757 & 35.1 \\
                \plants & 0.00109 & 44.450 & 0.99399 & 201.2 & 0.00076 & 46.290 & 0.99573 & 20.5 \\
                \roza   & 0.00123 & 44.068 & 0.99462 & 330.2 & 0.00080 & 48.261 & 0.99658 & 38.1 \\
                \dining & 0.00168 & 41.917 & 0.99066 & 300.0 & 0.00106 & 43.494 & 0.99280 & 40.1 \\
                \hline
                \multicolumn{9}{|c|}{Glossy}\\
                \hline
                \multirow{2}{*}{Scene} & \multicolumn{4}{c|}{\textit{k=64, l=4}} & \multicolumn{4}{c|}{\textit{k=128, l=4}}\\
                \cline{2-9}
                        & MAE     & PSNR   & SSIM    & FPS   & MAE     & PSNR   & SSIM    & FPS  \\
                \hline
                \teaset & 0.00133 & 45.752 & 0.99356 & 105.5 & 0.00102 & 48.129 & 0.99504 & 25.1 \\
                \plants & 0.00319 & 35.701 & 0.98237 &  64.5 & 0.00250 & 36.614 & 0.98661 & 12.2 \\
                \roza   & 0.00122 & 42.449 & 0.99701 & 120.5 & 0.00083 & 46.768 & 0.99817 & 31.1 \\
                \dining & 0.00411 & 32.638 & 0.97393 & 130.1 & 0.00312 & 33.414 & 0.97886 & 32.1 \\
                \hline
            \end{tabular}
            \caption{This table shows comparison among two different network sizes used to learn the transfer function. The quantitative metrics (MAE, PSNR \& SSIM) are calculated by comparing our rendering with baseline PRT as reference, which is a texture storage implementation of PRT. These metrics averaged amongst 30 uniformly sampled views in a trajectory for both \textsc{Diffuse} and \textsc{Glossy} material configurations. We also show the FPS obtained for respective network sizes. Our approach renders in real-time for both diffuse and glossy configurations.}
            \label{tab:mesh_network_comparisions}\vspace*{-3mm}
        \end{table*}

    \subsection{Results on SDF}
        \label{sub_sec:analytical_sdf_results}
        
        We now present our rendering results on scenes with SDF geometry. {\color{black}As stated earlier, implicit representations like SDFs do not have an inherent storage schema like the Mesh, which makes it very hard to store transfer vectors, hence we utilize the MLPs we trained to regress transfer.} Rendering is done using the GLSL implementation (Sect. \ref{sub_sec:glsl_implementation}). We sphere trace the SDF and obtain transfer at the intersection point using our network. Since sphere tracing needs to be done for all fragments, the scene geometry is set to two triangles that cover the entire image. We show results on four analytic SDFs, \mike, \rabbit from \cite{fizzler, quilez} and \oldcar, \fish from \cite{Takikawa2022SDF}. We also show results with one Neural SDF, \bunny using the method of \cite{single_nn_sdf}.
        
        Renderings with glossy materials using our method are shown in Fig. \ref{fig:analytical_sdf_glossy_results}. We render these scenes with two different environment lighting. All highlights from the environment map are reproduced on the glossy surface accurately. For instance, the \textit{yellowish} ground in the environment map reflects at the bottom of the SDF and the \textit{bluish} sky reflects at the top. We also show renderings of one diffuse and three glossy results with increasing Phong exponents on the \oldcar SDF in Fig. \ref{fig:teaser}. All scenes produce real-time FPS while producing plausible glossy highlights and soft shadows as shown in Tbl. \ref{tab:sdf_performance},  Figs. \ref{fig:analytical_sdf_glossy_results}, \ref{tab:sdf_performance}. The FPS for SDFs behaves similarly to triangle meshes, in that the $k=64, l=4$ network comfortably achieves real-time FPS while the $k=128, l=4$ network has a lower FPS.
        
        \begin{table}
            \centering
            \begin{tabular}{|c | c | c | c |c|}
                \hline
                \multirow{3}{*}{Scene} & \multicolumn{4}{c|}{FPS}\\
                \cline{2-5}
                                      &  \multicolumn{2}{c|}{\textit{k=64, l=4}}  & \multicolumn{2}{c|}{\textit{k=128, l=4}} \\
                \cline{2-5}
                                      & Diffuse   & Glossy                        & Diffuse          & Glossy                \\
                \hline
                \oldcar               & 254.96    & 207.28                        & 73.2             & 38.8                  \\
                
                \mike                 & 294.28    & 173.6                         & 61.0             & 24.2                  \\
                
                \fish                 & 320.1     & 231.6                         & 77.1             & 50.1                  \\
                
                \rabbit               & 310.21    & 205.94                        & 73.2             & 37.9                   \\
                \hline
            \end{tabular}
            \caption{Performance on diffuse and glossy renders of analytic SDFs. Our approach can render SDFs within the PRT framework in real time.}
            \label{tab:sdf_performance}\vspace*{-3mm}
        \end{table}
        
        \begin{figure*}
            \centering
            \begin{tikzpicture}
                \node at (-6.0, 2.) {\fish};
                \node at (-1.1, 2.) {\mike};
                \node at ( 3.8, 2.) {\rabbit};

                \node at (-6.0, 0.0) {\includegraphics[width=0.2\linewidth]{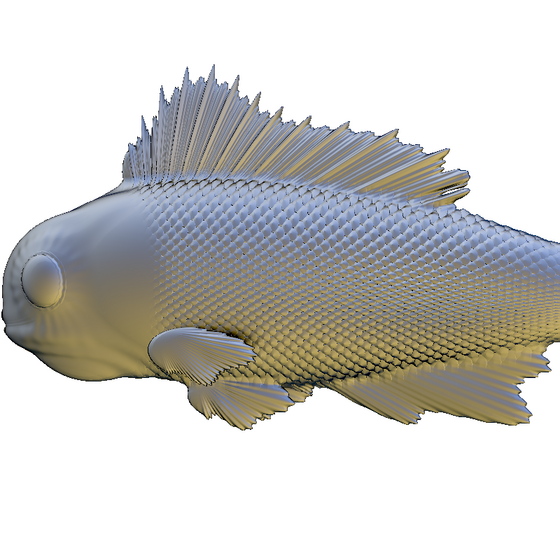}};
                \node at (-1.1, 0.0) {\includegraphics[width=0.2\linewidth]{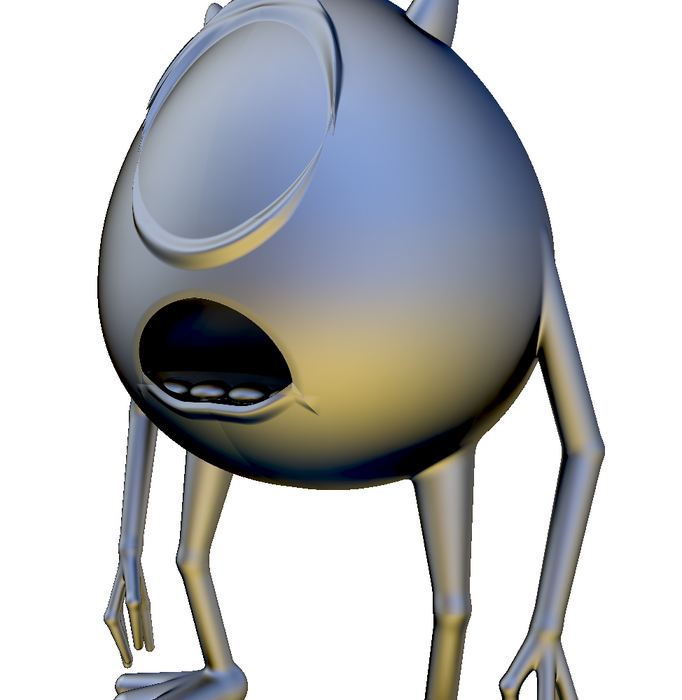}};
                \node at (3.8, 0.0) {\includegraphics[width=0.2\linewidth]{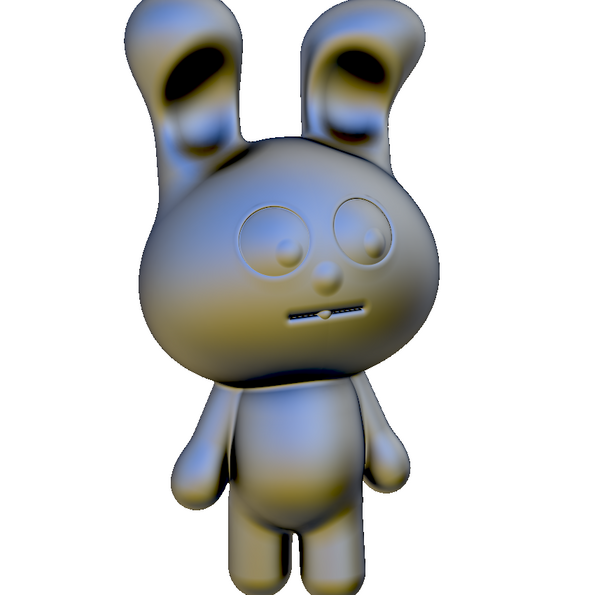}};
                \node at (-7.5, -1.7) {\includegraphics[width=0.1\linewidth]{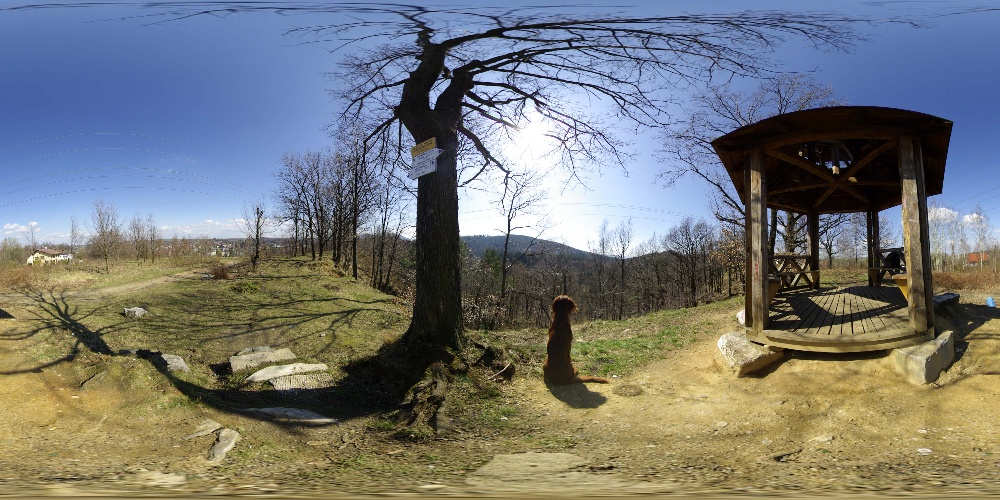}};
                \node at (-9.0, 0.0) {\rotatebox{90}{\textsc{Lighting 1}}};
                
                \node at (-6.0, -4.) {\includegraphics[width=0.2\linewidth]{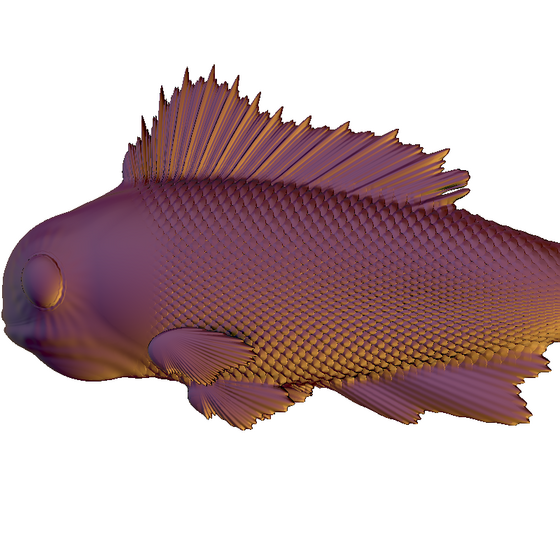}};
                \node at (-1.1, -4.) {\includegraphics[width=0.2\linewidth]{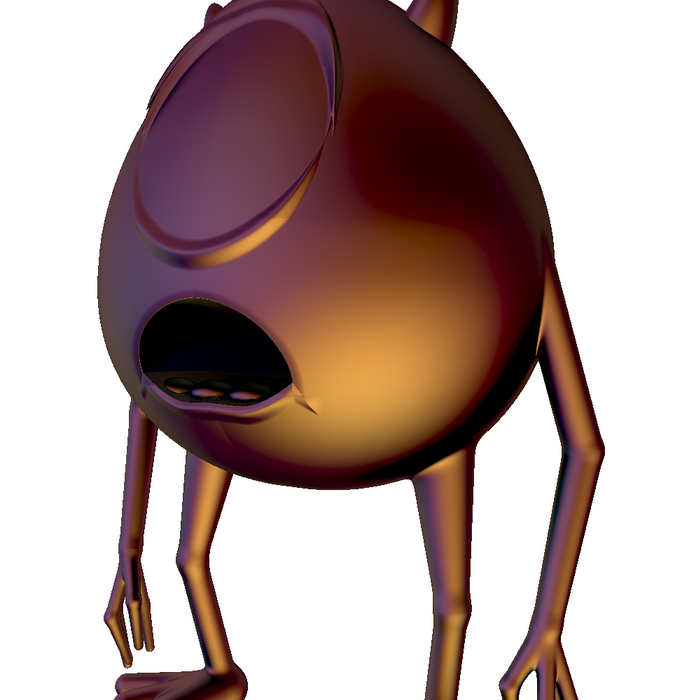}};
                \node at (3.8, -4.) {\includegraphics[width=0.2\linewidth]{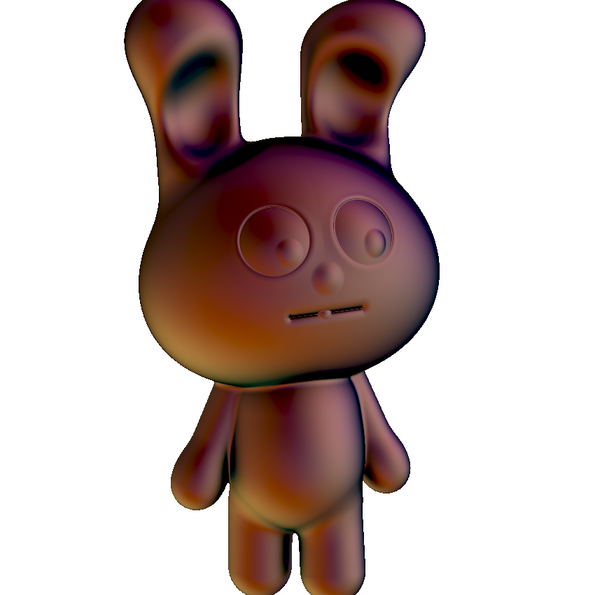}};
                \node at (-7.5, -5.7) {\includegraphics[width=0.1\linewidth]{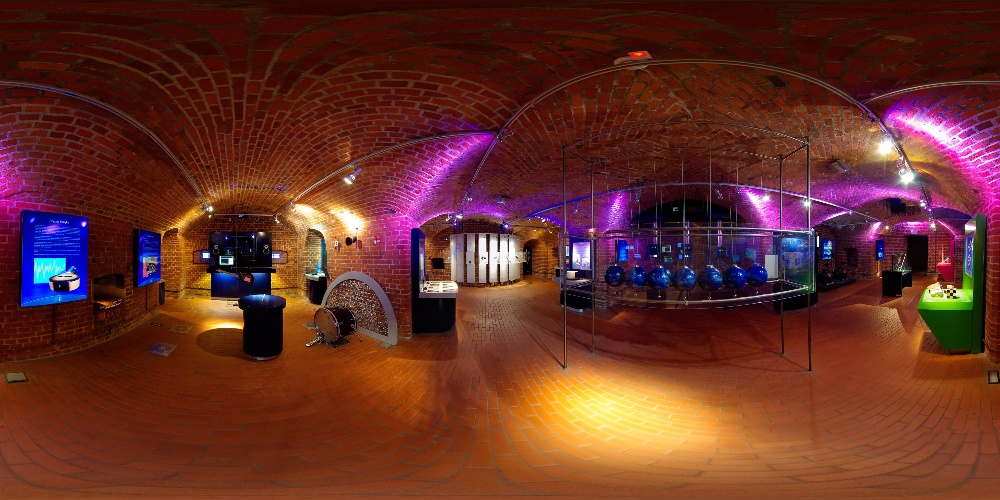}};
                \node at (-9.0, -4.) {\rotatebox{90}{\textsc{Lighting 2}}};
                
            \end{tikzpicture}
            \caption{SDF Results: With the use of learnt transfer approach we extended PRT onto SDF where UV mapping or vertex storage is not defined. The learnt function approximates the transfer well. Results are shown on three SDFs: \mike, \fish, \rabbit with two different lighting environments.}
            \label{fig:analytical_sdf_glossy_results}\vspace*{-3mm}
        \end{figure*}
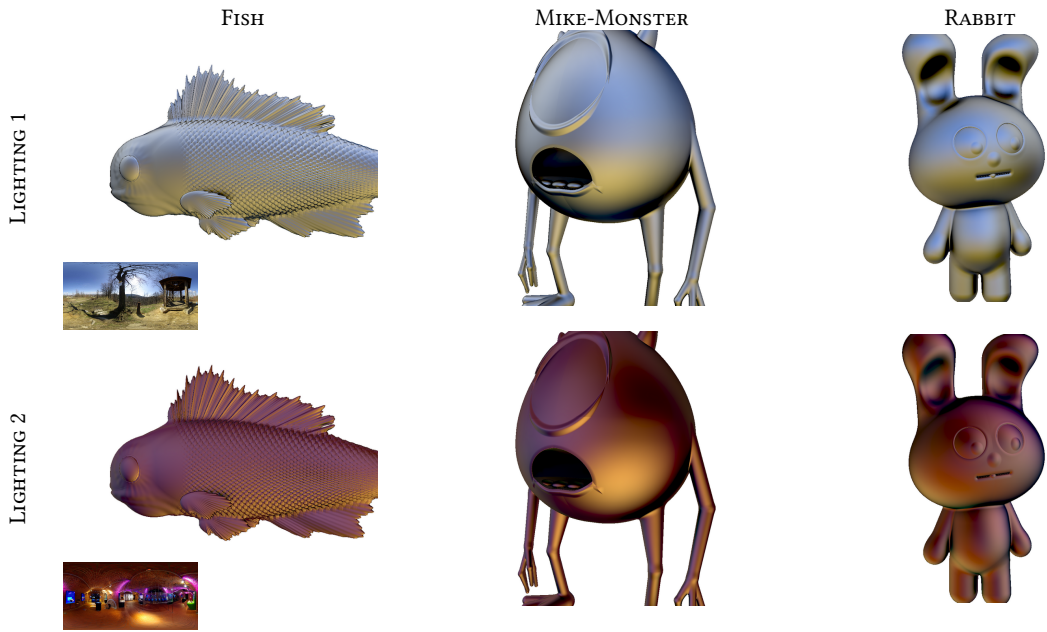
        
        \begin{figure}
            \centering
            \includegraphics[width=\linewidth]{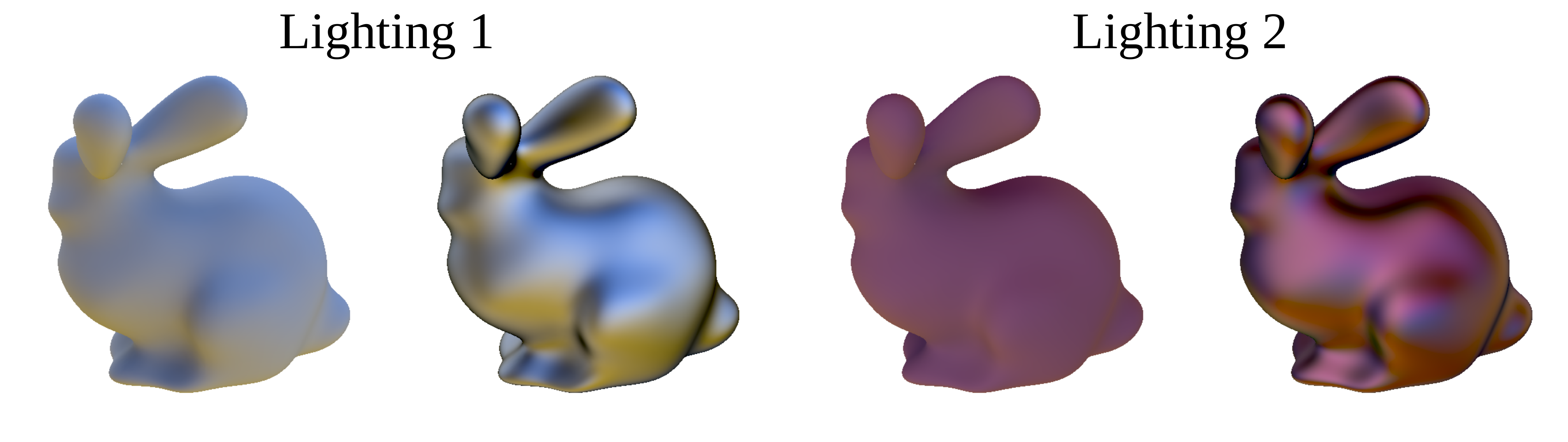}
            \caption{We show renderings of the \bunny neural SDF learnt using two network architectures. We show diffuse and glossy renderings of the neural SDF by evaluating our learnt transfer on the surface. Note that the geometric representation of \bunny is obtained from the \cite{single_nn_sdf} and we only contribute learnt transfer to it.}
            \label{fig:neural_sdf}\vspace*{-3mm}
        \end{figure}
        \begin{table}
            \centering
            \begin{tabular}{|c|c|c|c|c|}\hline
                \multirow{2}{*}{\backslashbox{Surface}{Transfer}} & \multicolumn{2}{c|}{\textit{k=64, l=4}} & \multicolumn{2}{c|}{\textit{k=128, l=4}} \\
                \cline{2-5}
                                                                  &  Diffuse      & Glossy                  & Diffuse     &    Glossy \\
                \hline
                \textit{k=32, l=2}                                &   320         & 200                     & 77          &    20.1   \\
                \hline
                \textit{k=64, l=2}                                &   50          & 25                      & 26.5         &    15.2   \\
                \hline
            \end{tabular}
            \caption{FPS on the \bunny Neural SDF (Fig. \ref{fig:neural_sdf}).}
            \label{tab:neural_sdf_performance}\vspace*{-3mm}
        \end{table}
        
        Finally, we show our rendering of the \bunny scene, which is defined as a Neural SDF as a proof of concept. We use the SDF optimized by \cite{single_nn_sdf}. The result is shown in Fig. \ref{fig:neural_sdf} for diffuse and glossy material and two different lighting conditions. We also show the FPS in Tbl. \ref{tab:neural_sdf_performance} for two different SDF networks sizes (vertical) and two different sizes of our network (horizontal). The renderings are plausible and achieve real-time framerates for smaller network sizes while maintaining interactivity for larger network sizes. The FPS could be further improved by optimizing the tracing of SDFs \cite{nglod}, a research direction orthogonal to ours.
        
        \begin{table}
            \centering
            \begin{tabular}{|c|c|c|c|c|}
            \hline
            Implementation & $k=64$  & $k=128$ & $k=256$ & $k=512$ \\ 
            \hline
            GLSL           & 64.4         & 12.2         & N/A          & N/A          \\ 
            \hline
            CUDA           & 55.1         & 29.1         & 20.7         & 8.0          \\
            \hline
            \end{tabular}
            \caption{CUDA vs GLSL: Table contains the Performance in FPS of the \glsl and \cuda implementation with various MLP architectures ranging from $k\in(64, 512), l=4$ tested on \plants scene with a glossy material. The \glsl implementation outperforms the \cuda implementation for $k=64$, but degrades at larger network sizes or fails to render ($k=256, k=512$). The \cuda implementation renders in all cases and is real-time for $k=64$ while maintaining interactivity for other cases. Please note we use the same number of layers($l=4$) in all the architectures presented in the table}
            \label{tab:cuda_vs_glsl_tab}\vspace*{-5mm}
        \end{table}
   
    \subsection{GLSL \& CUDA: Comparison}
        \label{sub_sec:cuda_vs_glsl}
        We compare and contrast the GLSL and CUDA implementations of our learnt representation. As mentioned before, the GLSL implementation limits the network size to the available local shader memory. On the other hand, CUDA allows the network to be as large as the entire GPU memory. We experimented with four different network sizes for both implementations: $k=64$, $k=128$, $k=256$ \& $k=512$ with $l=4$. The FPS for these configurations on the \plants scene are listed in Tbl. \ref{tab:cuda_vs_glsl_tab}. The GLSL implementation outperforms for $k=64$ network, while the performance drops on higher sizes. It can be observed that CUDA starts to take the lead from $k=128$ network. In fact, for $k=256$ and above, \glsl implementation fails to render. Although the performance of the CUDA implementation drops for higher network sizes, it is still able to load and render the scene interactively.\vspace*{-5mm}
        
    \subsection{Memory requirements}
        Thanks to our simple network structure, the entire transfer of a scene can be represented as a set of weights. In vertex-based storage of transfer as in \sloanprt, the memory requirement has a one-to-one correspondence with the mesh tessellation. With texture-based storage, this correspondence is reduced, with the caveat of needing a good UV-mapping or possibly a large texture. With our method, this correspondence is further reduced as the entire transfer is now just a set of weights. For example, a mesh with $1M$ vertices requires $\approx 70$MB data stored on vertices, while our method only needs around $64$KB - $5$MB. On the other-hand grid-based storage requires 256MB of data for a grid resolution of $512^3$.
}

\section{Discussion \& conclusion}
{
    \label{sec:discussion_conclusion}
    In this paper, we presented a learnt representation of transfer through small MLPs in traditional PRT frameworks. Our main motivation was to alleviate the dependence of transfer storage in PRT on geometry representations. We ensured that our network learns a \textit{continuous} representation of the transfer in a scene by densely and equally distributing training samples over the surface.  {\color{black}We provided a strategy to handle large-scenes in Sec.\ref{sub_sec:large_scenes} and Fig. \ref{fig:large_scenes} while maintaining real-time frame-rates.} We demonstrated two implementations of our learnt representation: GLSL and CUDA. We analyzed both implementations and discussed their merits and drawbacks. We further demonstrated that both implementations achieve real-time FPS on meshes as well as SDFs. SDF rendering which we demonstrated with our approach was not easily possible within the PRT framework. {\color{black}Utilizing our approach the works \cite{analytic_sh_many_poly_lights, analytic_sh_poly_light} can incorporate SDF geometries into their scenes.}

    One interesting challenge would be to empirically determine the optimal size of the network for a given complex scene, we would like to work on it in the future. We would also like to study the benefit of representing other surface properties, like specularity and albedo with our proposed approach. Finally, we would like to investigate the incorporation of inter-reflections with our approach, without baking in BRDF as done in \cite{sloan_2002}, for real-time global illumination with PRT with \textit{flexible surface representations}.
}

\noindent{\bf Acknowledgement: } Dhawal would like to thank \href{https://www.linkedin.com/in/pratikkumar-bulani-303302115/?originalSubdomain=in}{Pratikkumar Bulani} for helping set up the SDF scenes. Dhawal Sirikonda was partially supported by DST. Aakash KT was supported by the Kohli Research Fellowship. We also acknowledge support from the TCS Foundation through the Kohli Center on Intelligent Systems (KCIS).

\bibliographystyle{ACM-Reference-Format}
\bibliography{ICVGIP-Latex-Template}


\begin{thebibliography}{38}


\ifx \showCODEN    \undefined \def \showCODEN     #1{\unskip}     \fi
\ifx \showDOI      \undefined \def \showDOI       #1{#1}\fi
\ifx \showISBNx    \undefined \def \showISBNx     #1{\unskip}     \fi
\ifx \showISBNxiii \undefined \def \showISBNxiii  #1{\unskip}     \fi
\ifx \showISSN     \undefined \def \showISSN      #1{\unskip}     \fi
\ifx \showLCCN     \undefined \def \showLCCN      #1{\unskip}     \fi
\ifx \shownote     \undefined \def \shownote      #1{#1}          \fi
\ifx \showarticletitle \undefined \def \showarticletitle #1{#1}   \fi
\ifx \showURL      \undefined \def \showURL       {\relax}        \fi
\providecommand\bibfield[2]{#2}
\providecommand\bibinfo[2]{#2}
\providecommand\natexlab[1]{#1}
\providecommand\showeprint[2][]{arXiv:#2}

\bibitem[\protect\citeauthoryear{Davies, Nowrouzezahrai, and Jacobson}{Davies
  et~al\mbox{.}}{2021}]%
        {single_nn_sdf}
\bibfield{author}{\bibinfo{person}{Thomas~Ryan Davies}, \bibinfo{person}{Derek
  Nowrouzezahrai}, {and} \bibinfo{person}{Alec Jacobson}.}
  \bibinfo{year}{2021}\natexlab{}.
\newblock \showarticletitle{On the Effectiveness of Weight-Encoded Neural
  Implicit 3D Shapes}.
\newblock  (\bibinfo{year}{2021}).
\newblock


\bibitem[\protect\citeauthoryear{Dhawal, KT, and Narayanan}{Dhawal
  et~al\mbox{.}}{2022}]%
        {prtt}
\bibfield{author}{\bibinfo{person}{Sirikonda Dhawal}, \bibinfo{person}{Aakash
  KT}, {and} \bibinfo{person}{P.~J. Narayanan}.}
  \bibinfo{year}{2022}\natexlab{}.
\newblock \bibinfo{title}{PRTT: Precomputed Radiance Transfer Textures}.
\newblock
\newblock
\urldef\tempurl%
\url{https://doi.org/10.48550/ARXIV.2203.12399}
\showDOI{\tempurl}


\bibitem[\protect\citeauthoryear{Dombi}{Dombi}{2020}]%
        {moderngl}
\bibfield{author}{\bibinfo{person}{Szabolcs Dombi}.}
  \bibinfo{year}{2020}\natexlab{}.
\newblock \showarticletitle{ModernGL, high performance python bindings for
  OpenGL 3.3+}.
\newblock \bibinfo{publisher}{Github}.
\newblock


\bibitem[\protect\citeauthoryear{Fizzler}{Fizzler}{2018}]%
        {fizzler}
\bibfield{author}{\bibinfo{person}{Fizzler}.} \bibinfo{year}{2018}\natexlab{}.
\newblock \showarticletitle{Rabbit SH18}.
\newblock \bibinfo{publisher}{Shadertoy}.
\newblock
\urldef\tempurl%
\url{https://www.shadertoy.com/view/XlccWH}
\showURL{%
\tempurl}


\bibitem[\protect\citeauthoryear{Gafni, Thies, Zollh{\"{o}}fer, and
  Nie{\ss}ner}{Gafni et~al\mbox{.}}{2021}]%
        {gafani2021}
\bibfield{author}{\bibinfo{person}{Guy Gafni}, \bibinfo{person}{Justus Thies},
  \bibinfo{person}{Michael Zollh{\"{o}}fer}, {and} \bibinfo{person}{Matthias
  Nie{\ss}ner}.} \bibinfo{year}{2021}\natexlab{}.
\newblock \showarticletitle{Dynamic Neural Radiance Fields for Monocular 4D
  Facial Avatar Reconstruction}. In \bibinfo{booktitle}{\emph{{IEEE} Conference
  on Computer Vision and Pattern Recognition, {CVPR} 2021, virtual, June 19-25,
  2021}}. \bibinfo{publisher}{Computer Vision Foundation / {IEEE}},
  \bibinfo{pages}{8649--8658}.
\newblock
\urldef\tempurl%
\url{https://openaccess.thecvf.com/content/CVPR2021/html/Gafni\_Dynamic\_Neural\_Radiance\_Fields\_for\_Monocular\_4D\_Facial\_Avatar\_Reconstruction\_CVPR\_2021\_paper.html}
\showURL{%
\tempurl}


\bibitem[\protect\citeauthoryear{Greger, Shirley, Hubbard, and
  Greenberg}{Greger et~al\mbox{.}}{1998}]%
        {irradiance_vol}
\bibfield{author}{\bibinfo{person}{Gene Greger}, \bibinfo{person}{Peter
  Shirley}, \bibinfo{person}{Philip~M. Hubbard}, {and}
  \bibinfo{person}{Donald~P. Greenberg}.} \bibinfo{year}{1998}\natexlab{}.
\newblock \showarticletitle{The Irradiance Volume}.
\newblock \bibinfo{journal}{\emph{IEEE Comput. Graph. Appl.}}
  \bibinfo{volume}{18}, \bibinfo{number}{2} (\bibinfo{date}{mar}
  \bibinfo{year}{1998}), \bibinfo{pages}{32–43}.
\newblock
\showISSN{0272-1716}
\urldef\tempurl%
\url{https://doi.org/10.1109/38.656788}
\showDOI{\tempurl}


\bibitem[\protect\citeauthoryear{Hornik, Stinchcombe, and White}{Hornik
  et~al\mbox{.}}{1989}]%
        {universal_funct_approx}
\bibfield{author}{\bibinfo{person}{Kurt Hornik}, \bibinfo{person}{Maxwell
  Stinchcombe}, {and} \bibinfo{person}{Halbert White}.}
  \bibinfo{year}{1989}\natexlab{}.
\newblock \showarticletitle{Multilayer feedforward networks are universal
  approximators}.
\newblock \bibinfo{journal}{\emph{Neural Networks}} \bibinfo{volume}{2},
  \bibinfo{number}{5} (\bibinfo{year}{1989}), \bibinfo{pages}{359--366}.
\newblock
\showISSN{0893-6080}
\urldef\tempurl%
\url{https://doi.org/10.1016/0893-6080(89)90020-8}
\showDOI{\tempurl}


\bibitem[\protect\citeauthoryear{Iwanicki and Sloan}{Iwanicki and
  Sloan}{2009}]%
        {iwanicki}
\bibfield{author}{\bibinfo{person}{Michal Iwanicki} {and}
  \bibinfo{person}{Peter-Pike Sloan}.} \bibinfo{year}{2009}\natexlab{}.
\newblock \showarticletitle{Normal Mapping with Low-Frequency Precomputed
  Visibility}. In \bibinfo{booktitle}{\emph{SIGGRAPH 2009: Talks}} (New
  Orleans, Louisiana) \emph{(\bibinfo{series}{SIGGRAPH '09})}.
  \bibinfo{publisher}{ACM}, \bibinfo{address}{New York, NY, USA}, Article
  \bibinfo{articleno}{52}, \bibinfo{numpages}{1}~pages.
\newblock
\showISBNx{9781605588346}
\urldef\tempurl%
\url{https://doi.org/10.1145/1597990.1598042}
\showURL{%
\tempurl}


\bibitem[\protect\citeauthoryear{Kajiya}{Kajiya}{1986}]%
        {rendering_equation}
\bibfield{author}{\bibinfo{person}{James~T. Kajiya}.}
  \bibinfo{year}{1986}\natexlab{}.
\newblock \showarticletitle{The Rendering Equation}. In
  \bibinfo{booktitle}{\emph{Proceedings of the 13th Annual Conference on
  Computer Graphics and Interactive Techniques}}
  \emph{(\bibinfo{series}{SIGGRAPH '86})}. \bibinfo{publisher}{Association for
  Computing Machinery}, \bibinfo{address}{New York, NY, USA},
  \bibinfo{pages}{143–150}.
\newblock
\showISBNx{0897911962}
\urldef\tempurl%
\url{https://doi.org/10.1145/15922.15902}
\showDOI{\tempurl}


\bibitem[\protect\citeauthoryear{{Kl{\"o}ckner}, {Pinto}, {Lee}, {Catanzaro},
  {Ivanov}, and {Fasih}}{{Kl{\"o}ckner} et~al\mbox{.}}{2012}]%
        {kloeckner_pycuda_2012}
\bibfield{author}{\bibinfo{person}{Andreas {Kl{\"o}ckner}},
  \bibinfo{person}{Nicolas {Pinto}}, \bibinfo{person}{Yunsup {Lee}},
  \bibinfo{person}{B. {Catanzaro}}, \bibinfo{person}{Paul {Ivanov}}, {and}
  \bibinfo{person}{Ahmed {Fasih}}.} \bibinfo{year}{2012}\natexlab{}.
\newblock \showarticletitle{{PyCUDA and PyOpenCL: A Scripting-Based Approach to
  GPU Run-Time Code Generation}}.
\newblock \bibinfo{journal}{\emph{Parallel Comput.}} \bibinfo{volume}{38},
  \bibinfo{number}{3} (\bibinfo{year}{2012}), \bibinfo{pages}{157--174}.
\newblock
\showISSN{0167-8191}
\urldef\tempurl%
\url{https://doi.org/10.1016/j.parco.2011.09.001}
\showDOI{\tempurl}


\bibitem[\protect\citeauthoryear{K\v{r}iv\'{a}nek, Pattanaik, and
  \v{Z}\'{a}ra}{K\v{r}iv\'{a}nek et~al\mbox{.}}{2004}]%
        {adaptivemeshsubdiv2004}
\bibfield{author}{\bibinfo{person}{Jaroslav K\v{r}iv\'{a}nek},
  \bibinfo{person}{Sumanta Pattanaik}, {and} \bibinfo{person}{Ji\v{r}\'{\i}
  \v{Z}\'{a}ra}.} \bibinfo{year}{2004}\natexlab{}.
\newblock \showarticletitle{Adaptive Mesh Subdivision for Precomputed Radiance
  Transfer}. In \bibinfo{booktitle}{\emph{Proceedings of the 20th Spring
  Conference on Computer Graphics}} (Budmerice, Slovakia)
  \emph{(\bibinfo{series}{SCCG '04})}. \bibinfo{publisher}{Association for
  Computing Machinery}, \bibinfo{address}{New York, NY, USA},
  \bibinfo{pages}{106–111}.
\newblock
\showISBNx{1581139675}
\urldef\tempurl%
\url{https://doi.org/10.1145/1037210.1037226}
\showDOI{\tempurl}


\bibitem[\protect\citeauthoryear{Lehtinen, Zwicker, Turquin, Kontkanen, Durand,
  Sillion, and Aila}{Lehtinen et~al\mbox{.}}{2008}]%
        {lehtinen_hierarchical}
\bibfield{author}{\bibinfo{person}{Jaakko Lehtinen}, \bibinfo{person}{Matthias
  Zwicker}, \bibinfo{person}{Emmanuel Turquin}, \bibinfo{person}{Janne
  Kontkanen}, \bibinfo{person}{Fr\'{e}do Durand},
  \bibinfo{person}{Fran\c{c}ois~X. Sillion}, {and} \bibinfo{person}{Timo
  Aila}.} \bibinfo{year}{2008}\natexlab{}.
\newblock \showarticletitle{A Meshless Hierarchical Representation for Light
  Transport}. In \bibinfo{booktitle}{\emph{ACM SIGGRAPH 2008 Papers}} (Los
  Angeles, California) \emph{(\bibinfo{series}{SIGGRAPH '08})}.
  \bibinfo{publisher}{Association for Computing Machinery},
  \bibinfo{address}{New York, NY, USA}, Article \bibinfo{articleno}{37},
  \bibinfo{numpages}{9}~pages.
\newblock
\showISBNx{9781450301121}
\urldef\tempurl%
\url{https://doi.org/10.1145/1399504.1360636}
\showDOI{\tempurl}


\bibitem[\protect\citeauthoryear{Li, Wiedemann, and Mitchell}{Li
  et~al\mbox{.}}{2019}]%
        {deepdynamic_prt}
\bibfield{author}{\bibinfo{person}{Yue Li}, \bibinfo{person}{Pablo Wiedemann},
  {and} \bibinfo{person}{Kenny Mitchell}.} \bibinfo{year}{2019}\natexlab{}.
\newblock \showarticletitle{Deep Precomputed Radiance Transfer for Deformable
  Objects}.
\newblock \bibinfo{journal}{\emph{Proc. ACM Comput. Graph. Interact. Tech.}}
  \bibinfo{volume}{2}, \bibinfo{number}{1}, Article \bibinfo{articleno}{3}
  (\bibinfo{date}{jun} \bibinfo{year}{2019}), \bibinfo{numpages}{16}~pages.
\newblock
\urldef\tempurl%
\url{https://doi.org/10.1145/3320284}
\showDOI{\tempurl}


\bibitem[\protect\citeauthoryear{Lyu, Tewari, Leimkuehler, Habermann, and
  Theobalt}{Lyu et~al\mbox{.}}{2022}]%
        {nrtf}
\bibfield{author}{\bibinfo{person}{Linjie Lyu}, \bibinfo{person}{Ayush Tewari},
  \bibinfo{person}{Thomas Leimkuehler}, \bibinfo{person}{Marc Habermann}, {and}
  \bibinfo{person}{Christian Theobalt}.} \bibinfo{year}{2022}\natexlab{}.
\newblock \showarticletitle{Neural Radiance Transfer Fields for Relightable
  Novel-view Synthesis with Global Illumination}. In
  \bibinfo{booktitle}{\emph{European Conference on Computer Vision (ECCV)}}.
\newblock


\bibitem[\protect\citeauthoryear{Maas}{Maas}{2013}]%
        {Maas2013RectifierNI}
\bibfield{author}{\bibinfo{person}{Andrew~L. Maas}.}
  \bibinfo{year}{2013}\natexlab{}.
\newblock \showarticletitle{Rectifier Nonlinearities Improve Neural Network
  Acoustic Models}.
\newblock


\bibitem[\protect\citeauthoryear{McKenzie~Chapter}{McKenzie~Chapter}{2010}]%
        {mckenzie2010textured}
\bibfield{author}{\bibinfo{person}{Harrison~Lee McKenzie~Chapter}.}
  \bibinfo{year}{2010}\natexlab{}.
\newblock \showarticletitle{Textured Hierarchical Precomputed Radiance
  Transfer}.
\newblock  (\bibinfo{year}{2010}).
\newblock
\urldef\tempurl%
\url{https://doi.org/10.15368/theses.2010.101}
\showDOI{\tempurl}


\bibitem[\protect\citeauthoryear{Mildenhall, Srinivasan, Tancik, Barron,
  Ramamoorthi, and Ng}{Mildenhall et~al\mbox{.}}{2020}]%
        {mildenhall2020nerf}
\bibfield{author}{\bibinfo{person}{Ben Mildenhall}, \bibinfo{person}{Pratul~P.
  Srinivasan}, \bibinfo{person}{Matthew Tancik}, \bibinfo{person}{Jonathan~T.
  Barron}, \bibinfo{person}{Ravi Ramamoorthi}, {and} \bibinfo{person}{Ren Ng}.}
  \bibinfo{year}{2020}\natexlab{}.
\newblock \showarticletitle{NeRF: Representing Scenes as Neural Radiance Fields
  for View Synthesis}. In \bibinfo{booktitle}{\emph{ECCV}}.
\newblock


\bibitem[\protect\citeauthoryear{Ng, Ramamoorthi, and Hanrahan}{Ng
  et~al\mbox{.}}{2004}]%
        {triple_ren}
\bibfield{author}{\bibinfo{person}{Ren Ng}, \bibinfo{person}{Ravi Ramamoorthi},
  {and} \bibinfo{person}{Pat Hanrahan}.} \bibinfo{year}{2004}\natexlab{}.
\newblock \showarticletitle{Triple Product Wavelet Integrals for All-Frequency
  Relighting}.
\newblock \bibinfo{journal}{\emph{ACM Trans. Graph.}} \bibinfo{volume}{23},
  \bibinfo{number}{3} (\bibinfo{date}{Aug.} \bibinfo{year}{2004}),
  \bibinfo{pages}{477–487}.
\newblock
\showISSN{0730-0301}
\urldef\tempurl%
\url{https://doi.org/10.1145/1015706.1015749}
\showDOI{\tempurl}


\bibitem[\protect\citeauthoryear{Pantaleoni, Fascione, Hill, and
  Aila}{Pantaleoni et~al\mbox{.}}{2010}]%
        {prt_movie}
\bibfield{author}{\bibinfo{person}{Jacopo Pantaleoni}, \bibinfo{person}{Luca
  Fascione}, \bibinfo{person}{Martin Hill}, {and} \bibinfo{person}{Timo Aila}.}
  \bibinfo{year}{2010}\natexlab{}.
\newblock \showarticletitle{PantaRay: Fast Ray-Traced Occlusion Caching of
  Massive Scenes}.
\newblock \bibinfo{journal}{\emph{ACM Trans. Graph.}} \bibinfo{volume}{29},
  \bibinfo{number}{4}, Article \bibinfo{articleno}{37} (\bibinfo{date}{July}
  \bibinfo{year}{2010}), \bibinfo{numpages}{10}~pages.
\newblock
\urldef\tempurl%
\url{https://doi.org/10.1145/1778765.1778774}
\showDOI{\tempurl}


\bibitem[\protect\citeauthoryear{Paszke, Gross, Massa, Lerer, Bradbury, Chanan,
  Killeen, Lin, Gimelshein, Antiga, Desmaison, Kopf, Yang, DeVito, Raison,
  Tejani, Chilamkurthy, Steiner, Fang, Bai, and Chintala}{Paszke
  et~al\mbox{.}}{2019}]%
        {pytorch}
\bibfield{author}{\bibinfo{person}{Adam Paszke}, \bibinfo{person}{Sam Gross},
  \bibinfo{person}{Francisco Massa}, \bibinfo{person}{Adam Lerer},
  \bibinfo{person}{James Bradbury}, \bibinfo{person}{Gregory Chanan},
  \bibinfo{person}{Trevor Killeen}, \bibinfo{person}{Zeming Lin},
  \bibinfo{person}{Natalia Gimelshein}, \bibinfo{person}{Luca Antiga},
  \bibinfo{person}{Alban Desmaison}, \bibinfo{person}{Andreas Kopf},
  \bibinfo{person}{Edward Yang}, \bibinfo{person}{Zachary DeVito},
  \bibinfo{person}{Martin Raison}, \bibinfo{person}{Alykhan Tejani},
  \bibinfo{person}{Sasank Chilamkurthy}, \bibinfo{person}{Benoit Steiner},
  \bibinfo{person}{Lu Fang}, \bibinfo{person}{Junjie Bai}, {and}
  \bibinfo{person}{Soumith Chintala}.} \bibinfo{year}{2019}\natexlab{}.
\newblock \showarticletitle{PyTorch: An Imperative Style, High-Performance Deep
  Learning Library}.
\newblock In \bibinfo{booktitle}{\emph{Advances in Neural Information
  Processing Systems 32}}, \bibfield{editor}{\bibinfo{person}{H.~Wallach},
  \bibinfo{person}{H.~Larochelle}, \bibinfo{person}{A.~Beygelzimer},
  \bibinfo{person}{F.~d\textquotesingle Alch\'{e}-Buc},
  \bibinfo{person}{E.~Fox}, {and} \bibinfo{person}{R.~Garnett}} (Eds.).
  \bibinfo{publisher}{Curran Associates, Inc.}, \bibinfo{pages}{8024--8035}.
\newblock


\bibitem[\protect\citeauthoryear{Phong}{Phong}{1975}]%
        {phong}
\bibfield{author}{\bibinfo{person}{Bui~Tuong Phong}.}
  \bibinfo{year}{1975}\natexlab{}.
\newblock \showarticletitle{Illumination for Computer Generated Pictures}.
\newblock \bibinfo{journal}{\emph{Commun. ACM}} \bibinfo{volume}{18},
  \bibinfo{number}{6} (\bibinfo{date}{jun} \bibinfo{year}{1975}),
  \bibinfo{pages}{311–317}.
\newblock
\showISSN{0001-0782}
\urldef\tempurl%
\url{https://doi.org/10.1145/360825.360839}
\showDOI{\tempurl}


\bibitem[\protect\citeauthoryear{Quilez}{Quilez}{2013}]%
        {quilez}
\bibfield{author}{\bibinfo{person}{Inigo Quilez}.}
  \bibinfo{year}{2013}\natexlab{}.
\newblock \showarticletitle{Pixar Mike Monster Inc.}
\newblock \bibinfo{publisher}{Shadertoy}.
\newblock
\urldef\tempurl%
\url{https://www.shadertoy.com/view/MsXGWr}
\showURL{%
\tempurl}


\bibitem[\protect\citeauthoryear{Rahaman, Baratin, Arpit, Draxler, Lin,
  Hamprecht, Bengio, and Courville}{Rahaman et~al\mbox{.}}{2019}]%
        {pmlr-v97-rahaman19a}
\bibfield{author}{\bibinfo{person}{Nasim Rahaman}, \bibinfo{person}{Aristide
  Baratin}, \bibinfo{person}{Devansh Arpit}, \bibinfo{person}{Felix Draxler},
  \bibinfo{person}{Min Lin}, \bibinfo{person}{Fred Hamprecht},
  \bibinfo{person}{Yoshua Bengio}, {and} \bibinfo{person}{Aaron Courville}.}
  \bibinfo{year}{2019}\natexlab{}.
\newblock \showarticletitle{On the Spectral Bias of Neural Networks}. In
  \bibinfo{booktitle}{\emph{Proceedings of the 36th International Conference on
  Machine Learning}} \emph{(\bibinfo{series}{Proceedings of Machine Learning
  Research}, Vol.~\bibinfo{volume}{97})},
  \bibfield{editor}{\bibinfo{person}{Kamalika Chaudhuri} {and}
  \bibinfo{person}{Ruslan Salakhutdinov}} (Eds.). \bibinfo{publisher}{PMLR},
  \bibinfo{pages}{5301--5310}.
\newblock
\urldef\tempurl%
\url{https://proceedings.mlr.press/v97/rahaman19a.html}
\showURL{%
\tempurl}


\bibitem[\protect\citeauthoryear{Rainer, Bousseau, Ritschel, and
  Drettakis}{Rainer et~al\mbox{.}}{2022}]%
        {neural_prt}
\bibfield{author}{\bibinfo{person}{Gilles Rainer}, \bibinfo{person}{Adrien
  Bousseau}, \bibinfo{person}{Tobias Ritschel}, {and} \bibinfo{person}{George
  Drettakis}.} \bibinfo{year}{2022}\natexlab{}.
\newblock \showarticletitle{Neural Precomputed Radiance Transfer}.
\newblock \bibinfo{journal}{\emph{Computer Graphics Forum (Proceedings of the
  Eurographics conference)}} \bibinfo{volume}{41}, \bibinfo{number}{2}
  (\bibinfo{date}{April} \bibinfo{year}{2022}).
\newblock
\urldef\tempurl%
\url{http://www-sop.inria.fr/reves/Basilic/2022/RBRD22}
\showURL{%
\tempurl}


\bibitem[\protect\citeauthoryear{Ramamoorthi}{Ramamoorthi}{2009}]%
        {prt_ravir}
\bibfield{author}{\bibinfo{person}{Ravi Ramamoorthi}.}
  \bibinfo{year}{2009}\natexlab{}.
\newblock \bibinfo{booktitle}{\emph{Precomputation-Based Rendering}}.
\newblock \bibinfo{publisher}{NOW Publishers Inc}.
\newblock


\bibitem[\protect\citeauthoryear{Ramamoorthi and Hanrahan}{Ramamoorthi and
  Hanrahan}{2001}]%
        {irradiance_map}
\bibfield{author}{\bibinfo{person}{Ravi Ramamoorthi} {and} \bibinfo{person}{Pat
  Hanrahan}.} \bibinfo{year}{2001}\natexlab{}.
\newblock \showarticletitle{An Efficient Representation for Irradiance
  Environment Maps}. In \bibinfo{booktitle}{\emph{Proceedings of the 28th
  Annual Conference on Computer Graphics and Interactive Techniques}}
  \emph{(\bibinfo{series}{SIGGRAPH '01})}. \bibinfo{publisher}{Association for
  Computing Machinery}, \bibinfo{address}{New York, NY, USA},
  \bibinfo{pages}{497–500}.
\newblock
\showISBNx{158113374X}
\urldef\tempurl%
\url{https://doi.org/10.1145/383259.383317}
\showDOI{\tempurl}


\bibitem[\protect\citeauthoryear{Reiser, Peng, Liao, and Geiger}{Reiser
  et~al\mbox{.}}{2021}]%
        {Reiser2021ICCV}
\bibfield{author}{\bibinfo{person}{Christian Reiser}, \bibinfo{person}{Songyou
  Peng}, \bibinfo{person}{Yiyi Liao}, {and} \bibinfo{person}{Andreas Geiger}.}
  \bibinfo{year}{2021}\natexlab{}.
\newblock \showarticletitle{KiloNeRF: Speeding up Neural Radiance Fields with
  Thousands of Tiny MLPs}. In \bibinfo{booktitle}{\emph{International
  Conference on Computer Vision (ICCV)}}.
\newblock


\bibitem[\protect\citeauthoryear{Ren, Wang, Gong, Lin, Tong, and Guo}{Ren
  et~al\mbox{.}}{2013}]%
        {rrfs}
\bibfield{author}{\bibinfo{person}{Peiran Ren}, \bibinfo{person}{Jinpeng Wang},
  \bibinfo{person}{Minmin Gong}, \bibinfo{person}{Stephen Lin},
  \bibinfo{person}{Xin Tong}, {and} \bibinfo{person}{Baining Guo}.}
  \bibinfo{year}{2013}\natexlab{}.
\newblock \showarticletitle{Global Illumination with Radiance Regression
  Functions }.
\newblock \bibinfo{journal}{\emph{ACM Transactions on Graphics (TOG) - SIGGRAPH
  2013 Conference Proceedings}}  \bibinfo{volume}{32} (\bibinfo{date}{July}
  \bibinfo{year}{2013}).
\newblock
\urldef\tempurl%
\url{https://www.microsoft.com/en-us/research/publication/global-illumination-radiance-regression-functions/}
\showURL{%
\tempurl}


\bibitem[\protect\citeauthoryear{Sloan}{Sloan}{2008}]%
        {sloan2008stupid}
\bibfield{author}{\bibinfo{person}{Peter-Pike Sloan}.}
  \bibinfo{year}{2008}\natexlab{}.
\newblock \showarticletitle{Stupid spherical harmonics (sh) tricks}. In
  \bibinfo{booktitle}{\emph{Game developers conference}},
  Vol.~\bibinfo{volume}{9}. \bibinfo{pages}{42}.
\newblock


\bibitem[\protect\citeauthoryear{Sloan, Kautz, and Snyder}{Sloan
  et~al\mbox{.}}{2002}]%
        {sloan_2002}
\bibfield{author}{\bibinfo{person}{Peter-Pike Sloan}, \bibinfo{person}{Jan
  Kautz}, {and} \bibinfo{person}{John Snyder}.}
  \bibinfo{year}{2002}\natexlab{}.
\newblock \showarticletitle{Precomputed Radiance Transfer for Real-Time
  Rendering in Dynamic, Low-Frequency Lighting Environments}. In
  \bibinfo{booktitle}{\emph{Proceedings of the 29th Annual Conference on
  Computer Graphics and Interactive Techniques}} (San Antonio, Texas)
  \emph{(\bibinfo{series}{SIGGRAPH '02})}. \bibinfo{publisher}{Association for
  Computing Machinery}, \bibinfo{address}{New York, NY, USA},
  \bibinfo{pages}{527–536}.
\newblock
\showISBNx{1581135211}
\urldef\tempurl%
\url{https://doi.org/10.1145/566570.566612}
\showDOI{\tempurl}


\bibitem[\protect\citeauthoryear{Sloan, Hall, Hart, and Snyder}{Sloan
  et~al\mbox{.}}{2003}]%
        {pca_prt_sloan}
\bibfield{author}{\bibinfo{person}{Peter-Pike~J. Sloan}, \bibinfo{person}{J.
  Hall}, \bibinfo{person}{J. Hart}, {and} \bibinfo{person}{John~M. Snyder}.}
  \bibinfo{year}{2003}\natexlab{}.
\newblock \showarticletitle{Clustered principal components for precomputed
  radiance transfer}.
\newblock \bibinfo{journal}{\emph{ACM SIGGRAPH 2003 Papers}}
  (\bibinfo{year}{2003}).
\newblock


\bibitem[\protect\citeauthoryear{Takikawa, Glassner, and McGuire}{Takikawa
  et~al\mbox{.}}{2022}]%
        {Takikawa2022SDF}
\bibfield{author}{\bibinfo{person}{Towaki Takikawa}, \bibinfo{person}{Andrew
  Glassner}, {and} \bibinfo{person}{Morgan McGuire}.}
  \bibinfo{year}{2022}\natexlab{}.
\newblock \showarticletitle{A Dataset and Explorer for 3D Signed Distance
  Functions}.
\newblock \bibinfo{journal}{\emph{Journal of Computer Graphics Techniques
  (JCGT)}} \bibinfo{volume}{11}, \bibinfo{number}{2} (\bibinfo{date}{27 April}
  \bibinfo{year}{2022}), \bibinfo{pages}{1--29}.
\newblock
\showISSN{2331-7418}
\urldef\tempurl%
\url{http://jcgt.org/published/0011/02/01/}
\showURL{%
\tempurl}


\bibitem[\protect\citeauthoryear{Takikawa, Litalien, Yin, Kreis, Loop,
  Nowrouzezahrai, Jacobson, McGuire, and Fidler}{Takikawa
  et~al\mbox{.}}{2021}]%
        {nglod}
\bibfield{author}{\bibinfo{person}{Towaki Takikawa}, \bibinfo{person}{Joey
  Litalien}, \bibinfo{person}{Kangxue Yin}, \bibinfo{person}{Karsten Kreis},
  \bibinfo{person}{Charles Loop}, \bibinfo{person}{Derek Nowrouzezahrai},
  \bibinfo{person}{Alec Jacobson}, \bibinfo{person}{Morgan McGuire}, {and}
  \bibinfo{person}{Sanja Fidler}.} \bibinfo{year}{2021}\natexlab{}.
\newblock \showarticletitle{Neural Geometric Level of Detail: Real-time
  Rendering with Implicit {3D} Shapes}.
\newblock  (\bibinfo{year}{2021}).
\newblock


\bibitem[\protect\citeauthoryear{Tancik, Srinivasan, Mildenhall,
  Fridovich-Keil, Raghavan, Singhal, Ramamoorthi, Barron, and Ng}{Tancik
  et~al\mbox{.}}{2020}]%
        {tancik2020fourfeat}
\bibfield{author}{\bibinfo{person}{Matthew Tancik}, \bibinfo{person}{Pratul~P.
  Srinivasan}, \bibinfo{person}{Ben Mildenhall}, \bibinfo{person}{Sara
  Fridovich-Keil}, \bibinfo{person}{Nithin Raghavan}, \bibinfo{person}{Utkarsh
  Singhal}, \bibinfo{person}{Ravi Ramamoorthi}, \bibinfo{person}{Jonathan~T.
  Barron}, {and} \bibinfo{person}{Ren Ng}.} \bibinfo{year}{2020}\natexlab{}.
\newblock \showarticletitle{Fourier Features Let Networks Learn High Frequency
  Functions in Low Dimensional Domains}.
\newblock \bibinfo{journal}{\emph{NeurIPS}} (\bibinfo{year}{2020}).
\newblock


\bibitem[\protect\citeauthoryear{Turk}{Turk}{1990}]%
        {turk1990}
\bibfield{author}{\bibinfo{person}{Greg Turk}.}
  \bibinfo{year}{1990}\natexlab{}.
\newblock \bibinfo{booktitle}{\emph{Generating Random Points in Triangles}}.
\newblock \bibinfo{publisher}{Academic Press Professional, Inc.},
  \bibinfo{address}{USA}, \bibinfo{pages}{24–28}.
\newblock
\showISBNx{0122861695}


\bibitem[\protect\citeauthoryear{Wang and Ramamoorthi}{Wang and
  Ramamoorthi}{2018}]%
        {analytic_sh_poly_light}
\bibfield{author}{\bibinfo{person}{Jingwen Wang} {and} \bibinfo{person}{Ravi
  Ramamoorthi}.} \bibinfo{year}{2018}\natexlab{}.
\newblock \showarticletitle{Analytic Spherical Harmonic Coefficients for
  Polygonal Area Lights}.
\newblock \bibinfo{journal}{\emph{ACM Trans. Graph.}} \bibinfo{volume}{37},
  \bibinfo{number}{4}, Article \bibinfo{articleno}{54} (\bibinfo{date}{July}
  \bibinfo{year}{2018}), \bibinfo{numpages}{11}~pages.
\newblock
\showISSN{0730-0301}
\urldef\tempurl%
\url{https://doi.org/10.1145/3197517.3201291}
\showDOI{\tempurl}


\bibitem[\protect\citeauthoryear{Wu, Cai, Zhao, and Ramamoorthi}{Wu
  et~al\mbox{.}}{2020}]%
        {analytic_sh_many_poly_lights}
\bibfield{author}{\bibinfo{person}{Lifan Wu}, \bibinfo{person}{Guangyan Cai},
  \bibinfo{person}{Shuang Zhao}, {and} \bibinfo{person}{Ravi Ramamoorthi}.}
  \bibinfo{year}{2020}\natexlab{}.
\newblock \showarticletitle{Analytic Spherical Harmonic Gradients for Real-Time
  Rendering with Many Polygonal Area Lights}.
\newblock \bibinfo{journal}{\emph{ACM Trans. Graph.}} \bibinfo{volume}{39},
  \bibinfo{number}{4}, Article \bibinfo{articleno}{134} (\bibinfo{date}{July}
  \bibinfo{year}{2020}), \bibinfo{numpages}{14}~pages.
\newblock
\showISSN{0730-0301}
\urldef\tempurl%
\url{https://doi.org/10.1145/3386569.3392373}
\showDOI{\tempurl}


\bibitem[\protect\citeauthoryear{Zhang, Srinivasan, Deng, Debevec, Freeman, and
  Barron}{Zhang et~al\mbox{.}}{2021}]%
        {nerfactor2021}
\bibfield{author}{\bibinfo{person}{Xiuming Zhang}, \bibinfo{person}{Pratul~P.
  Srinivasan}, \bibinfo{person}{Boyang Deng}, \bibinfo{person}{Paul Debevec},
  \bibinfo{person}{William~T. Freeman}, {and} \bibinfo{person}{Jonathan~T.
  Barron}.} \bibinfo{year}{2021}\natexlab{}.
\newblock \showarticletitle{NeRFactor: Neural Factorization of Shape and
  Reflectance under an Unknown Illumination}.
\newblock \bibinfo{journal}{\emph{ACM Trans. Graph.}} \bibinfo{volume}{40},
  \bibinfo{number}{6}, Article \bibinfo{articleno}{237} (\bibinfo{date}{dec}
  \bibinfo{year}{2021}), \bibinfo{numpages}{18}~pages.
\newblock
\showISSN{0730-0301}
\urldef\tempurl%
\url{https://doi.org/10.1145/3478513.3480496}
\showDOI{\tempurl}


\end{thebibliography}

\appendix




\end{document}